\documentclass[aps,twocolumn,showpacs,preprintnumbers,prl,superscriptaddress,groupedaddress,10pt]{revtex4-1}
\usepackage[utf8]{inputenc}
\usepackage{graphicx,amssymb,amsmath,amsthm,amsfonts,epsfig,times,natbib}

\usepackage[usenames,dvipsnames]{color}
\usepackage{epstopdf}
\usepackage{amsmath,amssymb}
\usepackage{tensor}
\usepackage{mathtools}
\usepackage{amsbsy}
\usepackage{bm,url}
\usepackage[linktocpage]{hyperref}

\usepackage[scaled]{beramono}
\usepackage[T1]{fontenc}

\definecolor{oxfordblue}{rgb}{0.0, 0.13, 0.28}
\definecolor{burgundy}{rgb}{0.5, 0.0, 0.13}
\definecolor{darkolivegreen}{rgb}{0.33, 0.42, 0.18}
\definecolor{darkblue}{rgb}{0,0,0.5}
\definecolor{richcarmine}{rgb}{0.84, 0.0, 0.25}
\definecolor{darkblue}{rgb}{0,0,0.5}
\definecolor{venetianred}{rgb}{0.78, 0.03, 0.08}
\definecolor{skobeloff}{rgb}{0.0, 0.48, 0.45}
\hypersetup{colorlinks=true, citecolor=darkblue, linkcolor=darkblue,
urlcolor = darkblue}

\newcommand{\ben}{\begin{enumerate}}
\newcommand{\een}{\end{enumerate}}

\def\be{\begin{equation}}
\def\ee{\end{equation}}
\def\bea{\begin{eqnarray}}
\def\eea{\end{eqnarray}}

\newcommand{\beq}{\begin{eqnarray}}
\newcommand{\eeq}{\end{eqnarray}} 
\newcommand{\ba}{\begin{align}}
\newcommand{\ea}{\end{align}}

\begin{document}

\title{Spontaneously scalarised Kerr black holes}

\author{
Pedro V. P. Cunha$^{1,2}$,
Carlos A. R. Herdeiro$^{2}$,
Eugen Radu$^{1}$
}

 \affiliation{${^1}$ Departamento de F\'isica da Universidade de Aveiro and CIDMA,
Campus de Santiago, 3810-183 Aveiro, Portugal}
\affiliation{${^2}$ Centro de Astrofísica e Gravitação -- CENTRA,
Departamento de Física, Instituto Superior Técnico -- IST, Universidade de Lisboa -- UL,
Avenida Rovisco Pais 1, 1049-001 Lisboa, Portugal }
%


\date{\today}

\begin{abstract}
We construct asymptotically flat, spinning, regular on and outside an event horizon, scalarised black holes (SBHs) in extended scalar-tensor-Gauss-Bonnet models. They reduce to Kerr BHs when the scalar field vanishes. For an illustrative choice of non-minimal coupling, we scan the domain of existence. For each value of spin, SBHs exist in an interval between two critical masses, with the lowest one vanishing in the static limit. Non-uniqueness with Kerr BHs of equal global charges is observed; the SBHs are entropically favoured. This suggests SBHs  form dynamically from the spontaneous scalarisation of Kerr BHs, which are prone to a scalar-triggered tachyonic instability, below the largest critical mass. Phenomenologically, the introduction of BH spin damps the maximal observable difference between comparable scalarised and vacuum BHs. In the static limit, (perturbatively stable) SBHs can store over 20\% of the spacetime energy outside the event horizon; in comparison with Schwarzschild BHs, their geodesic frequency at the ISCO can differ by a factor of 2.5 and deviations in the shadow areal radius may top 40\%. As the BH spin grows, low mass SBHs are excluded, and the maximal relative differences decrease, becoming of order $\sim$ few \% for dimensionless spin $j\gtrsim 0.5$. This reveals a \textit{spin selection effect}: non-GR effects are only significant for low spin. We discuss if and how the recently measured shadow size of the M87 supermassive BH,  constrains the length scale of the Gauss-Bonnet coupling. 
\end{abstract}




\maketitle
{\bf Introduction.}
Strong gravity entered the precision era. The breakthroughs in gravitational wave astrophysics~\cite{Abbott:2016blz,LIGOScientific:2018mvr} and the unveiling of the first black hole (BH) shadow image~\cite{Akiyama:2019cqa,Akiyama:2019fyp,Akiyama:2019eap}, are probing the true nature of BH candidates. Thus, the hypothesis that astrophysical BHs when near equilbrium are well described by the Kerr metric~\cite{Kerr:1963ud}  -- the \textit{Kerr hypothesis} -- can be tested to a new level of accuracy~\cite{Berti:2015itd,Cardoso:2016ryw,Barack:2018yly}.

A primary concern in such testing is to assess degeneracy. That is, to what extent other viable BH models mimic the Kerr phenomenology. Three theoretical requirements for a viable BH model are: it should  (i)  arise in a consistent and well motivated (effective field) theory of gravity; (ii) have a dynamical formation mechanism; (iii) be (sufficiently) stable~\cite{Degollado:2018ypf}. 
Thus, a pressing theoretical task is to investigate the phenomenology of models obeying these criteria. 

One such family of models arises in the context of the \textit{BH spontaneous scalarisation} mechanism~\cite{Doneva:2017bvd,Silva:2017uqg,Antoniou:2017acq}, akin to the scalarisation of neutron stars occurring in scalar-tensor models~\cite{Damour:1993hw}. By considering \textit{extended}-scalar tensor Gauss-Bonnet models (eSTGB), which include a Gauss-Bonnet (GB) term,  vacuum General Relativity (GR) BHs -  the Kerr family - may scalarise. On the one hand, this is a sound class of models with second order field equations, avoiding Ostrogradsky instabilities~\cite{Ostrogradsky:1850fid}, and with high energy physics motivations~\cite{Zwiebach:1985uq}. On the other hand, the scalarisation phenomenon yields a formation mechanism, and, in the only case studied hitherto - the static spherical case~\cite{Antoniou:2017hxj,Minamitsuji:2018xde,Bakopoulos:2018nui,Motohashi:2018mql,Brihaye:2018grv,Macedo:2019sem,Doneva:2019vuh,Myung:2019wvb,Brihaye:2019kvj}  - the scalarised BHs (SBHs) may be perturbatively stable~\cite{Blazquez-Salcedo:2018jnn,Silva:2018qhn}. In this letter we report on eSTGB \textit{spinning} SBHs, which may form from the spontaneous scalarisation of Kerr BHs.  We show that introducing (the astrophysically relevant) spin downsizes the phenomenological effects of scalarisation. Using our results, we discuss how the M87 BH shadow measurement~\cite{Akiyama:2019eap} may constrain the length scale of the eSTGB spontaneous scalarisation models.

{\bf The model.}
Non-minimal couplings often allow circumventing BH no-scalar hair theorems~\cite{Sotiriou:2011dz,Herdeiro:2015waa}. 
A \textit{dynamical} scenario of BH spontaneous scalarisation relies on three key ingredients: $(a)$ one augments Einstein's GR with a new (real) scalar field degree of freedom, $\phi$, describing a spacetime varying coupling. The proposal that fundamental couplings vary in space and time is old, $e.g.$~\cite{Dirac:1938mt,Brans:1961sx}, and, in particular,  motivated by different attempts at grand unification theories and quantum gravity; $(b)$ one adds to GR a source of gravity, $\mathcal{I}$, which can trigger a \textit{repulsive} gravitational effect for BHs, and assumes some function $f(\phi)$ describes the coupling strength of $\mathcal{I}$. This class of models is described by the action:
\begin{eqnarray}  
\label{action}
S=
\frac{1}{16\pi}\int d^4x \sqrt{-g} \left[  R - 2\partial_\mu \phi \partial^\mu \phi
 +  f(\phi) \mathcal{I}   \right], 
\end{eqnarray}  
where $R$ is the Ricci scalar of the spacetime metric $g_{\mu\nu}$; $(c)$ one chooses $f(\phi)$ so that both GR and non-GR (\textit{scalarised}) exist in the model and the former may become unstable. 

The choice of $\mathcal{I}$ in $(b)$ could be, $e.g.$, the familiar electromagnetic Lagrangian $\mathcal{I}_{\rm EM}=-F_{\mu\nu}F^{\mu\nu}$~\cite{Herdeiro:2018wub} or, the choice herein, the Gauss-Bonnet curvature squared term, 
\begin{equation}
\mathcal{I}_{\rm GB}=\lambda^2(R^{\mu\nu\alpha\beta}R_{\mu\nu\alpha\beta}-4R_{\mu\nu}R^{\mu\nu} +R^2) \ ,
\end{equation}
where $\lambda$ is a constant length scale. Choosing $\mathcal{I}_{\rm EM}$, electro-vacuum GR BHs become unstable below a certain $M/Q$, $i.e.$, BH mass to charge ratio, determined by the choice of $f(\phi)$, and spontaneously scalarise. In this model, scalarisation was established dynamically~\cite{Herdeiro:2018wub,Myung:2018jvi,Fernandes:2019rez} and shown to be approximately conservative, for high $M/Q$.  Choosing $\mathcal{I}_{\rm GB}$, vacuum GR BHs become unstable against scalarisation below a certain $M/\lambda$~\cite{Doneva:2017bvd}. In both cases, it becomes dynamically favourable for BHs with sufficiently low mass compared to the length scale associated to the repulsive effect ($Q$ or $\lambda$) to excite the scalar field, varying the coupling strength of the repulsive term.

{\bf Coupling function and instability threshold.}
Appropriate couplings $f(\phi)$ obey $f'(\phi)|_{\phi=0}\equiv df/d\phi|_{\phi=0}=0$, so that the vacuum GR BHs are solutions of~\eqref{action}. Moreover, the latter are prone to a tachyonic instability triggered  by a scalar field perturbation if $\mathcal{I} \, d^2f/d\phi^2|_{\phi=0}>0$~\cite{Herdeiro:2018wub}. A choice of appropriate coupling, that yields entropically favoured, perturbatively stable, static,  spherical SBHs in the eSTGB model, is~\cite{Doneva:2017bvd}: 
\begin{eqnarray}  
\label{fu}
f(\phi)=\frac{1}{2\beta}(1-e^{-\beta\phi^2}) \ ,
\end{eqnarray}  
where $\beta>0$. Following~\cite{Doneva:2017bvd} we take $\beta=6$. For sufficiently high $\beta$ the properties of this particular choice are universal. For Schwarzschild BHs, $\mathcal{I}_{\rm GB}>0$, and both above conditions are met for~\eqref{fu}.

\begin{figure}[h!]
\begin{center} 
\includegraphics[width=0.49\textwidth]{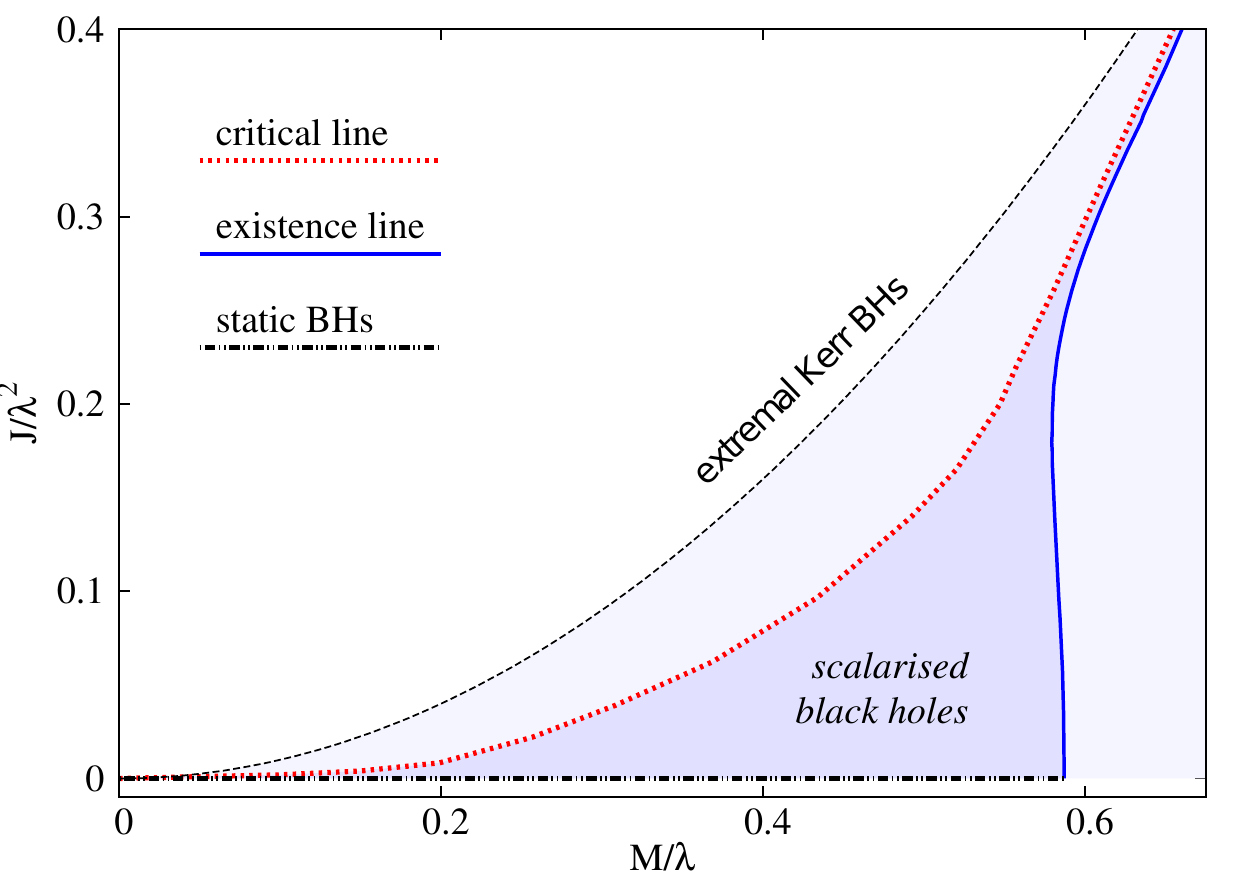}
%
\caption{ 
$(M,J)$-domain of existence of spinning SBHs.
}
\label{MJ}
\end{center}
\end{figure}

Schwarzschild BHs are unstable against scalarisation when~\cite{Doneva:2017bvd,Silva:2017uqg} $M/\lambda\lesssim 0.587$.
This number is independent of the specific choice of $f(\phi)$, as long as it is compatible with scalarisation. 
For a Kerr BH with given dimensionless spin $j\equiv J/M^2$, 
the corresponding threshold for stability is now a function of $J/\lambda^2$, forming an \textit{existence line} in the $(M,J)$-plane. We will see in Fig.~\ref{MJ} that there exist some $J$, such that the threshold of scalarisation is larger than $0.587$. Thus, spinning BHs can scalarise for (slightly) larger masses.

{\bf The domain of existence.}
Stationary, axisymmetric solutions of~\eqref{action} describing spinning, SBHs can be constructed using the metric ansatz $ds^2=-e^{2F_0} N dt^2+e^{2F_1}\left(dr^2/N +r^2 d\theta^2\right)+e^{2F_2}r^2 \sin^2\theta (d\varphi-W dt)^2$, 
where $N\equiv 1-r_H/r$,  and $F_i,W$, as well as the scalar field $\phi$, depend on $r,\theta$ only. Asymptotic flatness is guaranteed by imposing that 
$
\lim_{r\rightarrow \infty}{F_i}=\lim_{r\rightarrow \infty}{W}=\lim_{r\rightarrow \infty}{\phi}=0$.
 Axial symmetry and regularity impose, on the symmetry axis, $\theta=0,\pi$,
$
\partial_\theta F_i = \partial_\theta W = \partial_\theta \phi = 0
$,
and 
%
$
F_1=F_2 
$, (no conical singularities).
At the event horizon, located at a constant value of $r=r_H>0$, 
a new radial coordinate is convenient, 
$
x\equiv\sqrt{r^2-r_H^2}$, leading to the horizon boundary conditions, at $r=r_H$,
$
\partial_x F_i = \partial_x \phi  =  0$ and $W =\Omega_H$, 
where the constant $\Omega_H>0$ is the horizon angular velocity.

The system of coupled PDEs obtained from~\eqref{action} is solved using a Newton-Raphson relaxation method, the above boundary conditions and the same numerical strategy and solver as in, $e.g.$,~\cite{Herdeiro:2015gia}. The ADM mass $M$ and total angular momentum $J$ of the BH solutions are read off from 
the asymptotic expansions, for large $r$: $g_{tt} =-e^{2F_0}N+e^{2F_2}W^2r^2 \sin^2 \theta
\simeq-1+2M/r$, $g_{\varphi t}=-e^{2F_2}W r^2 \sin^2 \theta\simeq -2J\sin^2\theta/r$.  In Fig.~\ref{MJ}, we exhibit the domain of existence of the spinning SBHs in the $(M,J)$ plane (shaded blue region), obtained by extrapolating to the continuum a discrete set of thousands of solutions. 
 %
The domain of existence is bounded by the static BHs (black dashed line), which have $J=0$ and $M/\lambda<0.587$, the existence (blue solid) line corresponding to the Kerr BHs that can support test field configurations of the scalar field (a \textit{zero mode} of the instability) and a set of critical solutions (red dotted line). 
The existence line is universal for any $f(\phi)$ allowing scalarisation.
The critical and static sets are model dependent. The former reveals that below a certain $M/\lambda$, which depends on $J$, solutions cease to exist. This behaviour is shared with the dilaton-GB model~\cite{Kanti:1995vq,Kleihaus:2015aje}. Physically, the repulsive GB term can prevent the existence of an event horizon below a certain $M/\lambda$, and this value increases with $J$, which adds another repulsive effect~\footnote{Technically this limit can be understood from the perturbative expansion of the solutions near the horizon, wherein the radicand of a square root becomes negative.}.

\begin{figure}[h!]
\begin{center}
\includegraphics[width=0.49\textwidth]{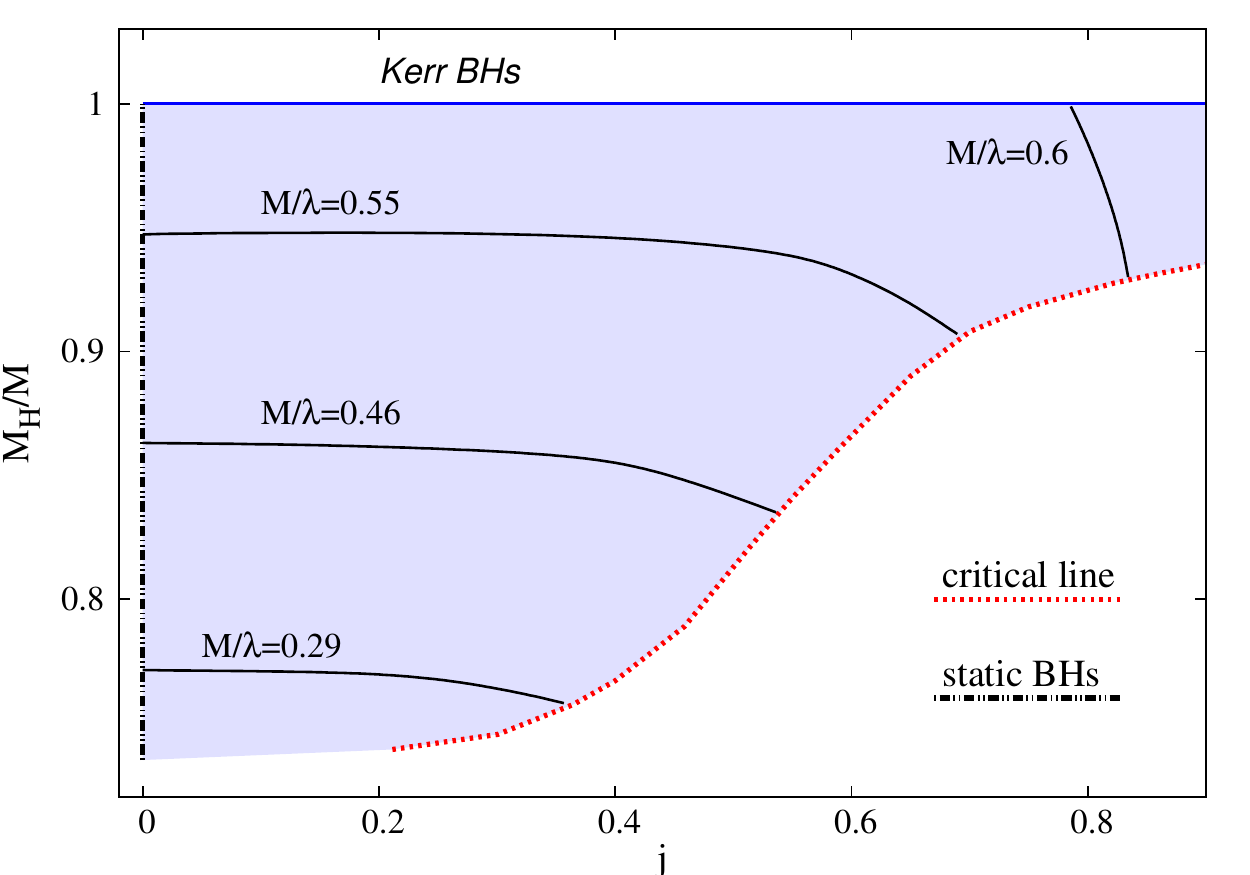}  
\caption{ 
Horizon mass to ADM mass ratio $vs.$ $j$. The colour coding in Figs.~\ref{MJ}-\ref{ae} is the same.
}
\label{MHM}
\end{center}
\end{figure}

{\bf Physical properties.}
Non-uniqueness between scalarised and Kerr BHs, for the same global quantities $(M,J)$, is manifest in Fig.~\ref{MJ}. To get a measure of how much these two families of BHs differ, we show in Fig.~\ref{MHM} the ratio of the horizon mass, computed as the Komar integral~\cite{Komar:1958wp} for the stationarity Killing vector field, $k=\partial_t$, to the ADM mass, as a function of the dimensionless spin. 
For static BHs ($j=0$) over 20\% of the spacetime energy can be stored outside the horizon. This occurs for solutions with the smallest values of $M/\lambda\lesssim 0.29$. For large spin, say $j\gtrsim 0.8$, solutions have $M/\lambda\gtrsim 0.6$ and the energy outside the horizon is less than 10\%.  In the $j=0$ limit, the SBHs in this model are stable against radial perturbations for $M/\lambda\gtrsim 0.171$~\cite{Blazquez-Salcedo:2018jnn}, which encompasses the whole static (black dashed) line in Fig.~\ref{MHM}.

\begin{figure}[h!]
\begin{center}
\includegraphics[width=0.49\textwidth]{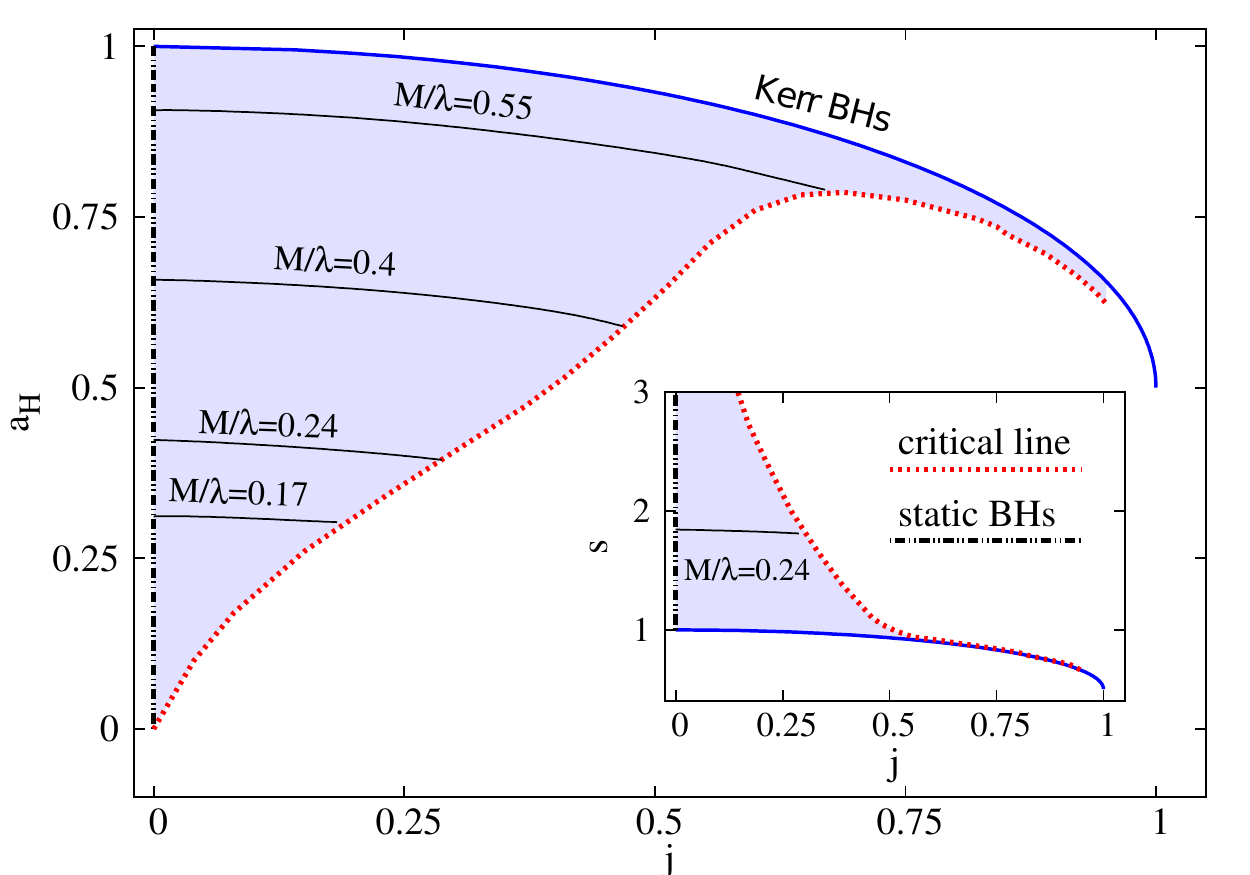}  
\caption{ 
Reduced area (main panel) and reduced entropy (inset) of the spinning SBHs $vs.$ $j$. 
}
\label{ae}
\end{center}
\end{figure}

Let us assess the entropy of the spinning SBHs. In the eSTGB model, the BH entropy is \textit{not} given by the Bekenstein-Hawking formula, $S_{\rm BH}=A_H/4$, where the event horizon area $A_H$, is, in terms of our metric ansatz, $A_H=2\pi r_H^2 \int_0^\pi d\theta \sin \theta~e^{F_1^{(2)}(\theta)+F_2^{(2)}(\theta)}$. The corrected entropy is obtained using Wald's approach~\cite{Wald:1993nt}, 
\begin{equation}
S=S_{\rm BH}+S_{sGB} \ ,  \ \ \  S_{sGB}\equiv \frac{\lambda^2}{2} \int_{H} d^2 x \sqrt{h}f(\phi) {\rm  R}^{(2)} \ ,
\end{equation} 
where ${\rm  R}^{(2)}$ is the Ricci scalar of the metric $h_{ij}$, induced on the spatial sections of the event horizon, $H$. Defining  the reduced (dimensionless) area and entropy,
\begin{equation}
a_H\equiv \frac{A_H}{16\pi M^2} \ , \qquad  s\equiv \frac{S}{4\pi M^2} \ ,
\end{equation} 
we plot these quantities in Fig.~\ref{ae} for the spinning SBHs. 
For fixed $j$, the reduced area of the SBHs decreases from the existence to the critical line. In this sense, the SBHs are smaller than comparable Kerr BHs (with the same $M,J$). But, by virtue of the corrected entropy formula, the reduced entropy of the spinning SBHs, for fixed $j$ increases from the existence to the critical line - $cf.$ Fig.~\ref{ae} (inset). Thus, SBHs are entropically preferred over comparable, $i.e.$ same $(M,J)$, Kerr BHs. 

The solutions obey a Smarr-type formula~\cite{Smarr:1972kt},
\begin{equation}
M=2( T_H S+\Omega_H J)+M_{(\phi)} \ ,
\end{equation}
where $T_H=(e^{F_0^{(2)}(\theta)-F_1^{(2)}(\theta)})/(4\pi r_H)$ is the Hawking temperature and $M_{(\phi)}$ is a bulk (outside the horizon) integral along a spacelike hypersurface $\Sigma$, accounting for the scalar field $
M_{(\phi)}=\frac{1}{2\pi} \int_\Sigma d^3 x \sqrt{-g} (f(\phi)/f'(\phi)) \Box \phi$. This formula is used for numerical accuracy tests. Another test is provided by the first law of BH thermodynamics, $
dM=T_H dS +\Omega_H dJ$, 
where no explicit scalar field term appears, even though the solutions possess a scalar ``charge" $Q_s$, which is read off from the far-field asymptotics as $\phi\simeq-Q_s/r$.

\begin{figure}[h!]
\begin{center}
\includegraphics[width=0.23\textwidth]{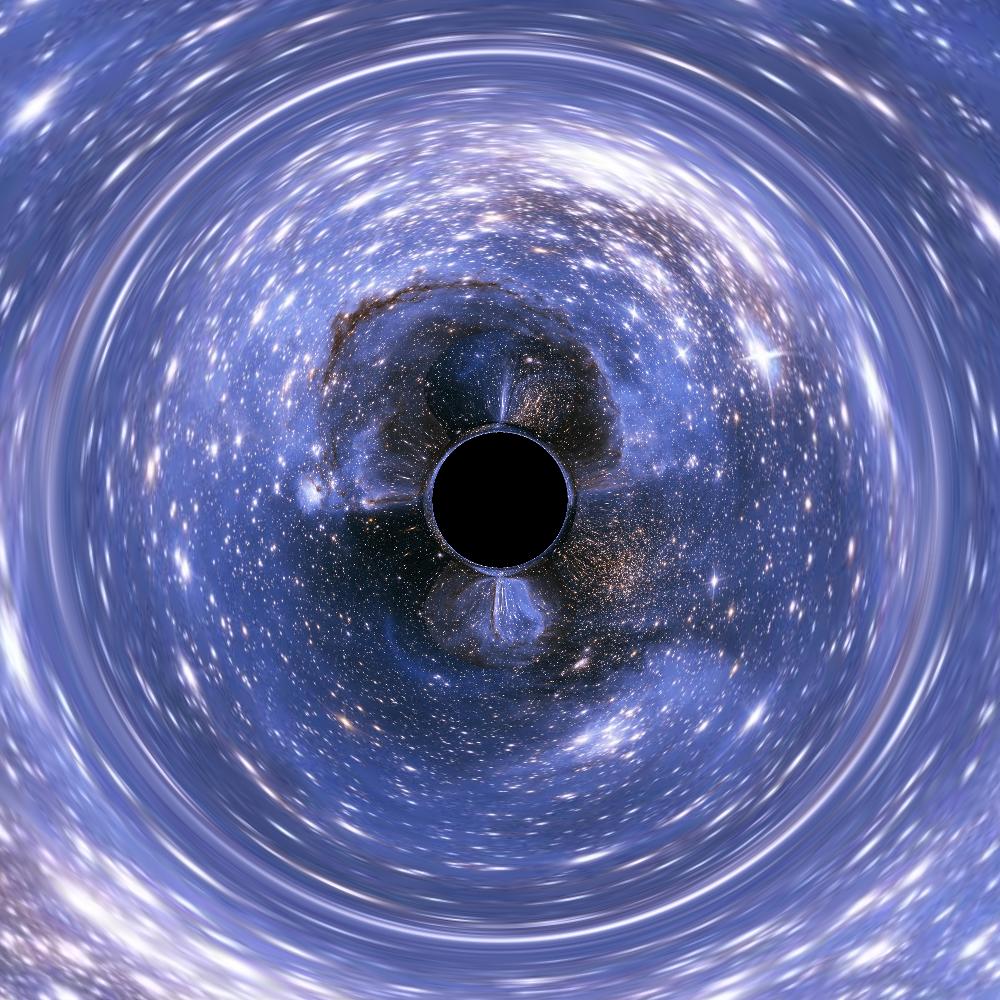} 
\includegraphics[width=0.23\textwidth]{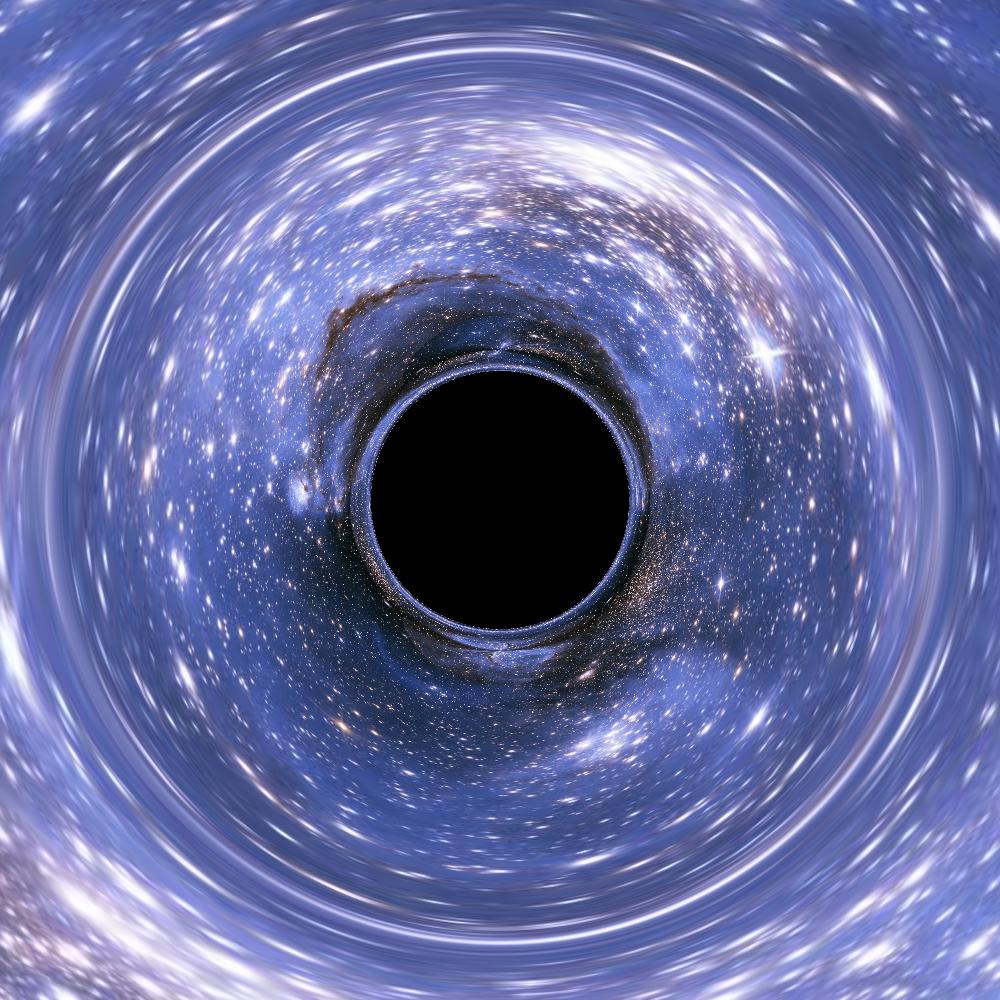}
\includegraphics[width=0.23\textwidth]{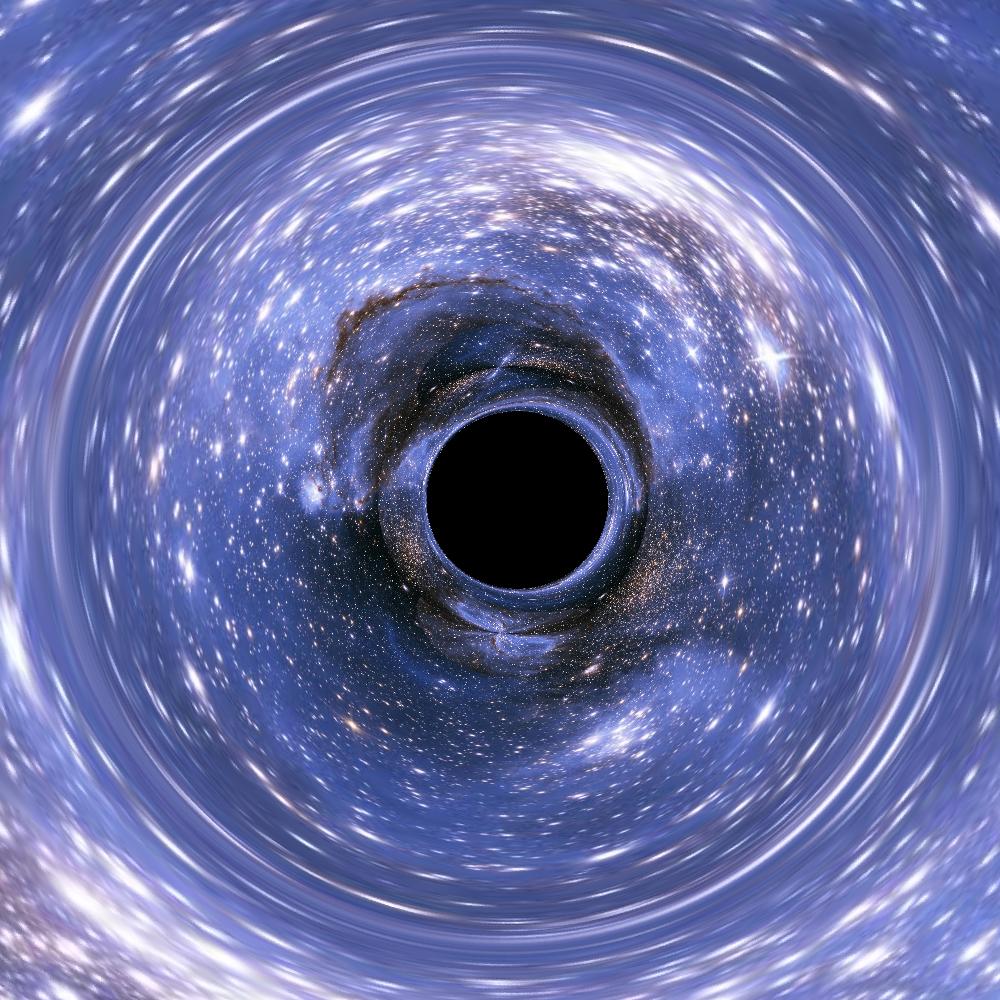}  
\includegraphics[width=0.23\textwidth]{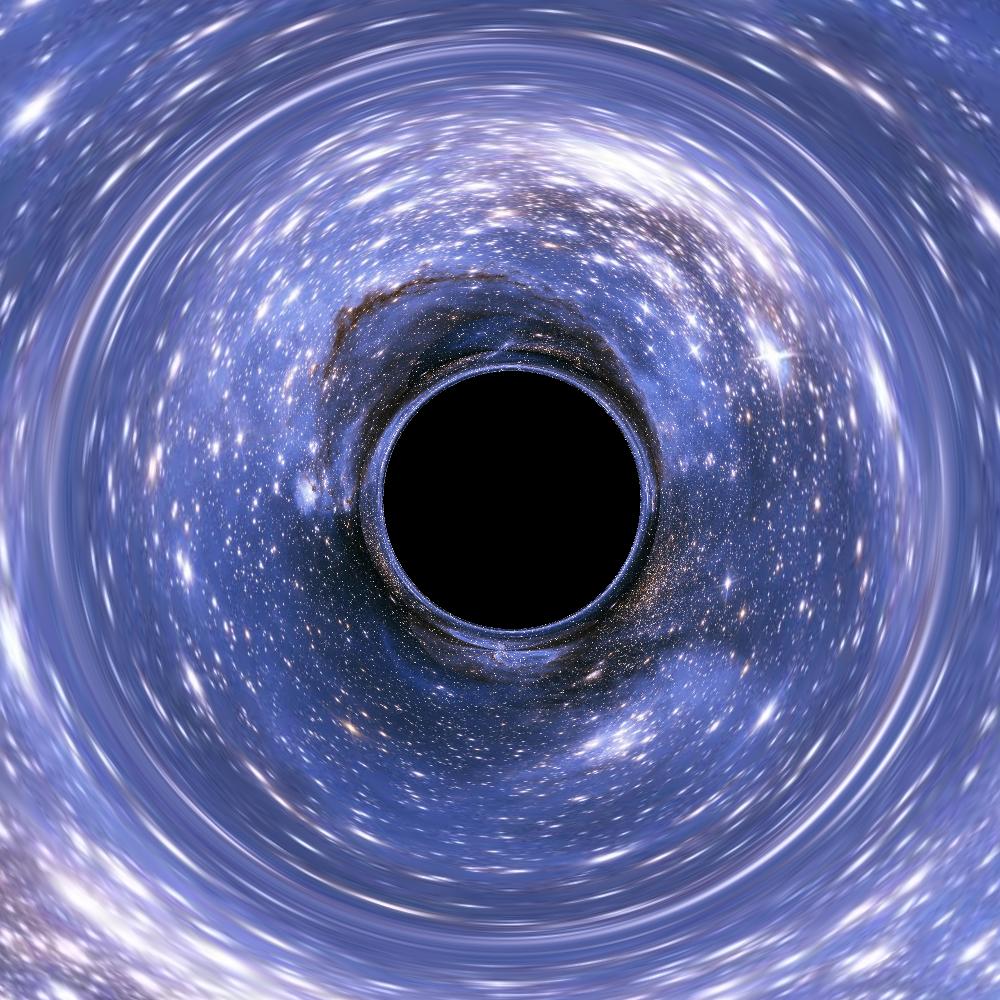}  

\caption{Shadows and lensing, under comparable observation conditions. (Top, left to right): (i) static SBH with $M/\lambda\sim 0.172$;  (ii) comparable Schwarzschild BH. (Bottom, left to right): (iii) spinning SBH with $M/\lambda\sim 0.237$ ($j=0.24$); (iv) comparable Kerr BH. The background image can be found in~\cite{image}.
}
\label{shadows1}
\end{center}
\end{figure}

\begin{figure}[h!]
\begin{center}
\includegraphics[width=0.49\textwidth]{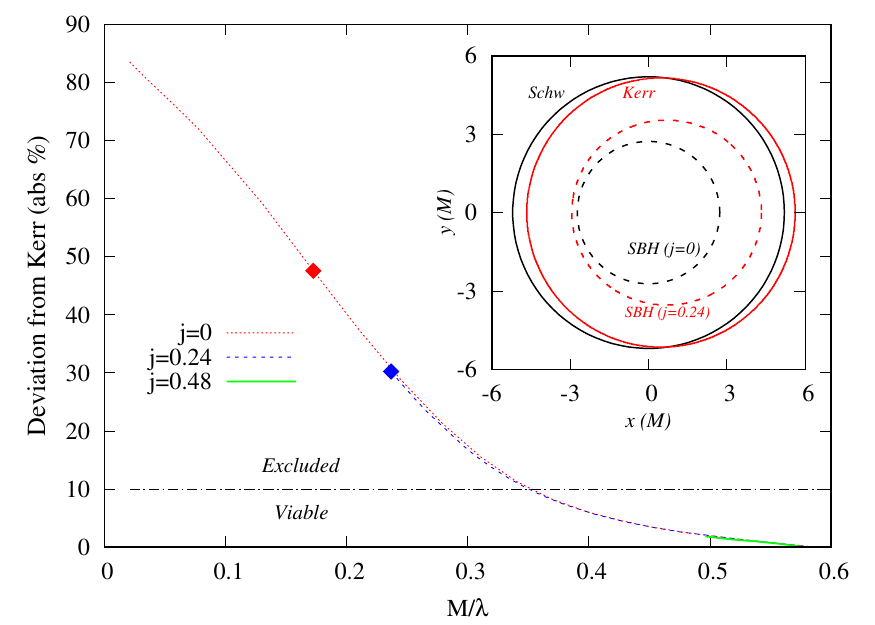}  
\caption{ 
Shadow size relative deviations  between SBHs and Kerr BHs. (Right inset) contours of the shadows in Fig.~\ref{shadows1} for comparison; the two SBHs are highlighted as diamonds in the main panel. 
}
\label{shadows2}
\end{center}
\end{figure}

{\bf BH shadows.}
The shadow is an optical image of a BH due to background or surrounding light sources~\cite{Perlick:2004tq,Cunha:2018acu,Cunha:2016wzk,Falcke:1999pj,Johannsen:2016vqy}. Recently the first image of a BH shadow was released~\cite{Akiyama:2019cqa,Akiyama:2019fyp,Akiyama:2019eap}. A BH shadow is a feature of strong gravitational lensing, and it is determined by the fundamental photon orbits - bound states of light around the BH~\cite{Cunha:2017eoe}. This include, in particular, the equatorial light rings (LRs). Measuring the LRs gives the boundary of the shadow.  The shadows in two distinct (stationary and axisymmetric) BH spacetimes are comparable if the BHs are comparable and the observers are identical, say, both on the equatorial plane and at the same perimetral distance $\sqrt{g_{\varphi\varphi}}$ from each BH~\cite{Cunha:2016bpi}. In Fig.~\ref{shadows1}  the shadow and lensing of two SBHs, and their comparable vacuum counterparts are shown, obtained using ray tracing. 
The $j=0$ SBH is the smallest stable one, in order to maximise the relative difference with its vacuum counterpart. The shadow of the former has roughly half the size, whereas the Einstein ring (the lensing of the image point behind the BH) is similar - top panels, Fig.~\ref{shadows1}. For the spinning SBH and Kerr BH (both with the spin value, $j=0.24$) the differences are still obvious, but smaller - lower panels, Fig.~\ref{shadows1}.  To make the comparison quantitative, we introduce the areal radius $\bar{r}\equiv\sqrt{A/\pi}$, for a shadow image with area $A$; $\bar{r}$ is well defined even for non circular shadows. This measure is used in Fig.~\ref{shadows2}, to obtain the relative deviation in the shadow size between SBHs and Kerr BHs. 
One observes the deviations increase monotonically as $M/\lambda$ decreases. For $j=0$, deviations can be larger than 80\%, or, restricting to the stable BHs ($M/\lambda\gtrsim 0.171$), larger than 40\%. This is consistent with SBHs being smaller, $cf.$ Fig.~\ref{ae}. Turning on $j$ introduces a critical lower mass, $cf.$ Fig.~\ref{MJ}, and the larger deviations (at the smallest $M/\lambda$) are absent. For $j=0.24$, deviations still reach $\sim$30\%, but for $j=0.48$, deviations are only of a few percent, resembling the Einstein-dilaton-GB model~\cite{Cunha:2016wzk}. The $j=0.24$ SBH shown in~Fig.~\ref{shadows1} maximises the difference with Kerr for this spin value.  For $j\gtrsim 0.5$ the shadows of SBHs differ from Kerr $\lesssim$ 2\% and lensing images would of this sort would look identical to the comparable Kerr.

{\bf The M87 shadow.}
The M87 supermassive BH imaging~\cite{Akiyama:2019cqa} gave a BH angular scale of $\theta_g= 3.8 \pm 0.4$ $\mu$as. We thus consider a $10\%$ error in the measured shadow size~\footnote{The corresponding Schwarzschild shadow \textit{diameter} is $6\sqrt{3}\,\theta_g\simeq 39$ $\mu$as. For a Kerr BH this value varies, increasing $j$, by $\lesssim$ 2\%, due to the spin inclination angle $\sim$ 17$^o$. There is an offset between such light ring diameter and the  \textit{measured} emission ring of $42\pm3$ $\mu$as;  $\theta_g$ is inferred from the latter via GRMHD models~\cite{Akiyama:2019eap} which predict a peak emission slightly outside the light ring.}.
%
%
A prior measurement of the BH mass using stellar dynamics~\cite{Gebhardt:2011yw}, but with updated distance~\cite{Akiyama:2019cqa}, gave $M_{\rm M87}=6.2^{+1.1}_{-0.6}\times 10^9 \, M_\odot$. Taking this as the true mass, a putative SBH is only consistent with the data (which is consistent with a Kerr BH) if the relative shadow deviation is $\lesssim$ 10\%. The corresponding line separating excluded from viable SBHs is shown in Fig.~\ref{shadows2}. Its intersection with the line of solutions is not very sensitive to $j$: $M/\lambda=0.35$ $(0.353)$ for $j=0$ $(j=0.24)$. Such low spin values are compatible with some estimates of the M87 BH spin, say, $j\sim 0.1$ in~\cite{Nokhrina:2019sxv}. Imposing $M_{\rm M87}/\lambda\gtrsim 0.353$, yields the (weak) constraint $\lambda \lesssim 1.8\times 10^{10} \, M_\odot$.
For larger $\lambda$, a BH with $M=M_{\rm M87}$ would scalarise, and, assuming this process to be approximately conservative~\footnote{This is supported by the qualitatively similar dynamical processes studied in the cousin model~\cite{Herdeiro:2018wub}.}, the SBH shadow would be too small. Even though the mass estimate in~\cite{Gebhardt:2011yw} assumes GR,  the GB term fall off as $\sim 1/r^6$ implies GR is a good approximation at $\sim$few gravitational radii in our model and thus in the study of stellar dynamics. 
On the other hand, if the M87 BH spin is large, as in other estimates, say, $j\sim 0.9$ in~\cite{Tamburini:2019vrf}, its shadow measurement~\cite{Akiyama:2019cqa} \textit{per se} is compatible with a SBH, as differences are below the 10\% error bar.

\begin{figure}[h!]
\begin{center}
\includegraphics[width=0.49\textwidth]{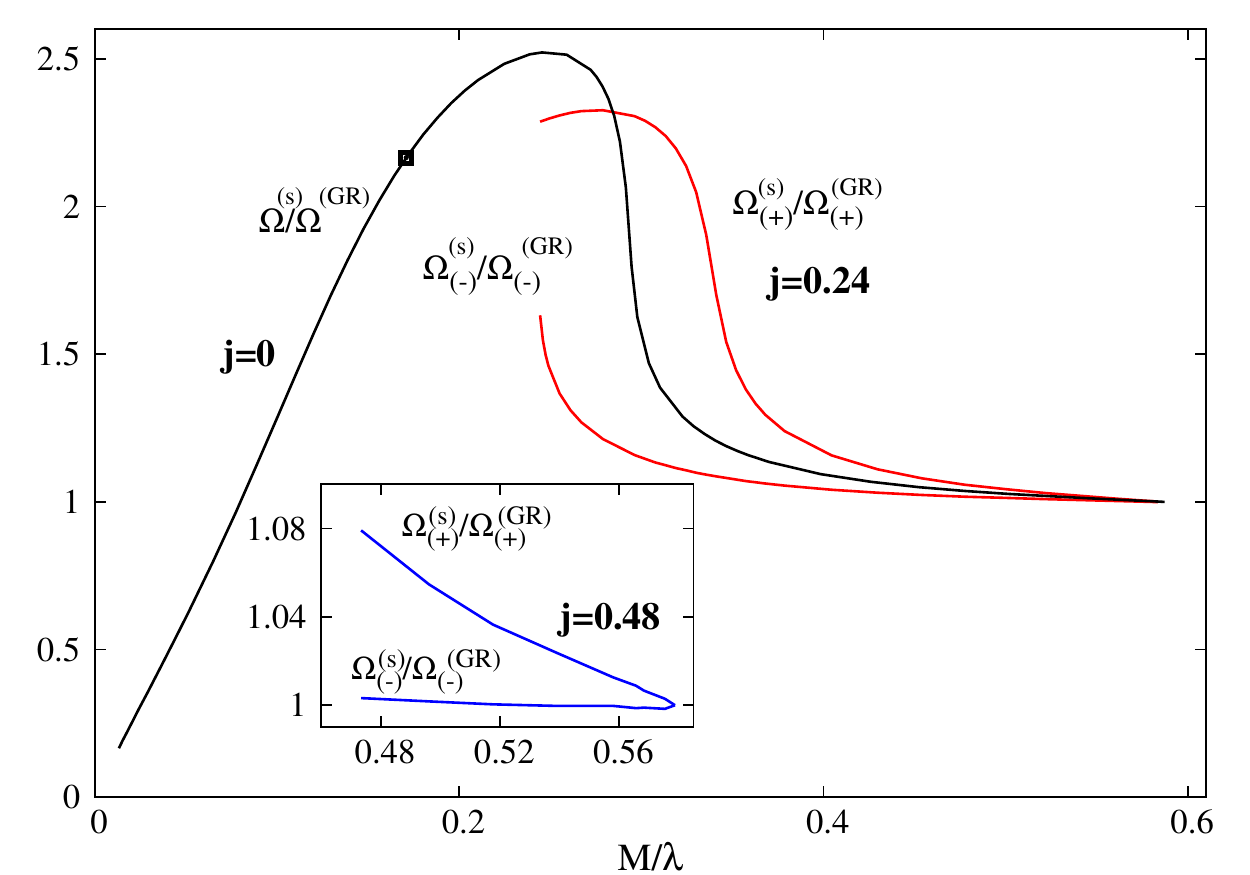}  
\caption{ 
Relative frequencies at the ISCO for scalarised and GR BHs.
}
\label{isco}
\end{center}
\end{figure}

{\bf Frequency at the ISCO.}
The geodesic frequency at the innermost stable circular orbit (ISCO) of the SBHs, $\Omega$, is of relevance for X-ray astronomy, since the inner edge of accretion disks is often modelled at the ISCO ($e.g.$ Thorne-Novikov disk model~\cite{1973blho.conf..343N}). Then, this frequency dictates the cut-off on the frequency of synchroton radiation, used in the continuum fitting of the X-ray spectrum of accreting BH systems~\cite{Bambi:2017khi}. The ISCO frequency computation follows, $e.g.$,~\cite{Herdeiro:2015gia}. Fig.~\ref{isco} shows the relative frequencies at the ISCO for SBHs and Kerr BHs. For $j=0$, the ratio $\Omega^{(s)}/\Omega^{(GR)}$ reaches a maximum $\sim$2.5, attained for stable BHs. The transition to unstable BHs is marked with a square ($M/\lambda\lesssim 0.171$). In the spinning case, for $j=0.24$, deviations are larger for prograde orbits [denoted with $(+)$] with a maximum ratio $\sim 2.3$. For $j=0.48$ (inset) deviation are $\lesssim$ 10\%, and even less for higher $j$. For $j=0$, $r_{ISCO}$ and $r_{LR}$ have a different sensitivity to $M/\lambda$. Varying $M/\lambda$ from $0.587\rightarrow 0.3$, the former approximately halves, whereas the latter becomes only $\sim$2/3 of its initial value. This explains the larger deviation from Schwarzschild in the ISCO behaviour as compared to shadows.

{\bf Universality and a spin selection effect.}
%
Other coupling functions $f(\phi)$, compatible with spontaneous scalarisation, $e.g.$~\cite{Silva:2017uqg}, imply a different range for $M/\lambda$ even in the static ($j=0$) case. We anticipate the impact of $j\neq 0$ should generically follow the trend herein: increase the
minimal value of $M/\lambda$ and simultaneously decreasing the maximal relative deviations from Kerr, as in~\cite{Cunha:2016wzk}.
This contrasts with the effect of spin in some models of scalarised stars~\cite{Doneva:2018ouu}.

Finally, we observe this model illustrates a \textit{spin selection effect}: BHs with moderate to large spins will be either Kerr BHs or SBHs which are undistinguishable from Kerr BHs in current observations. Only low spin BHs could clearly unveil the scalarisation phenomenon. In this regard, one may recall that typical BHs observed by both gravitational wave observations and X-ray spectroscopy methods are estimated to have a fairly large spin - $cf.$ Table III in~\cite{LIGOScientific:2018mvr} and Table 8.1 in~\cite{Bambi:2017khi}.



\bigskip

{\bf Acknowledgements.}
P.C.  is  supported  by
Grant No.  PD/BD/114071/2015 under the FCT-IDPASC Portugal Ph.D. program. This work is supported by the Funda\c{c}\~ao para a Ci\^encia e a Tecnologia (FCT) project UID/MAT/04106/2019 (CIDMA), by CENTRA (FCT) strategic project UID/FIS/00099/2013, by national funds (OE), through FCT, I.P., in the scope of the framework contract foreseen in the numbers 4, 5 and 6
of the article 23, of the Decree-Law 57/2016, of August 29,
changed by Law 57/2017, of July 19. We acknowledge support  from the project PTDC/FIS-OUT/28407/2017.   This work has further been supported by  the  European  Union's  Horizon  2020  research  and  innovation  (RISE) programmes H2020-MSCA-RISE-2015
Grant No.~StronGrHEP-690904 and H2020-MSCA-RISE-2017 Grant No.~FunFiCO-777740. The authors would like to acknowledge networking support by the
COST Action CA16104.

\bigskip


\bibliography{lettersk}

\begin{thebibliography}{57}%
\makeatletter
\providecommand \@ifxundefined [1]{%
 \@ifx{#1\undefined}
}%
\providecommand \@ifnum [1]{%
 \ifnum #1\expandafter \@firstoftwo
 \else \expandafter \@secondoftwo
 \fi
}%
\providecommand \@ifx [1]{%
 \ifx #1\expandafter \@firstoftwo
 \else \expandafter \@secondoftwo
 \fi
}%
\providecommand \natexlab [1]{#1}%
\providecommand \enquote  [1]{``#1''}%
\providecommand \bibnamefont  [1]{#1}%
\providecommand \bibfnamefont [1]{#1}%
\providecommand \citenamefont [1]{#1}%
\providecommand \href@noop [0]{\@secondoftwo}%
\providecommand \href [0]{\begingroup \@sanitize@url \@href}%
\providecommand \@href[1]{\@@startlink{#1}\@@href}%
\providecommand \@@href[1]{\endgroup#1\@@endlink}%
\providecommand \@sanitize@url [0]{\catcode `\\12\catcode `\$12\catcode
  `\&12\catcode `\#12\catcode `\^12\catcode `\_12\catcode `\%12\relax}%
\providecommand \@@startlink[1]{}%
\providecommand \@@endlink[0]{}%
\providecommand \url  [0]{\begingroup\@sanitize@url \@url }%
\providecommand \@url [1]{\endgroup\@href {#1}{\urlprefix }}%
\providecommand \urlprefix  [0]{URL }%
\providecommand \Eprint [0]{\href }%
\providecommand \doibase [0]{http://dx.doi.org/}%
\providecommand \selectlanguage [0]{\@gobble}%
\providecommand \bibinfo  [0]{\@secondoftwo}%
\providecommand \bibfield  [0]{\@secondoftwo}%
\providecommand \translation [1]{[#1]}%
\providecommand \BibitemOpen [0]{}%
\providecommand \bibitemStop [0]{}%
\providecommand \bibitemNoStop [0]{.\EOS\space}%
\providecommand \EOS [0]{\spacefactor3000\relax}%
\providecommand \BibitemShut  [1]{\csname bibitem#1\endcsname}%
\let\auto@bib@innerbib\@empty
\bibitem [{\citenamefont {Abbott}\ \emph {et~al.}(2016)\citenamefont {Abbott}
  \emph {et~al.}}]{Abbott:2016blz}%
  \BibitemOpen
  \bibfield  {author} {\bibinfo {author} {\bibfnamefont {B.~P.}\ \bibnamefont
  {Abbott}} \emph {et~al.} (\bibinfo {collaboration} {Virgo, LIGO
  Scientific}),\ }\href {\doibase 10.1103/PhysRevLett.116.061102} {\bibfield
  {journal} {\bibinfo  {journal} {Phys. Rev. Lett.}\ }\textbf {\bibinfo
  {volume} {116}},\ \bibinfo {pages} {061102} (\bibinfo {year} {2016})},\
  \Eprint {http://arxiv.org/abs/1602.03837} {arXiv:1602.03837 [gr-qc]}
  \BibitemShut {NoStop}%
\bibitem [{\citenamefont {Abbott}\ \emph {et~al.}(2018)\citenamefont {Abbott}
  \emph {et~al.}}]{LIGOScientific:2018mvr}%
  \BibitemOpen
  \bibfield  {author} {\bibinfo {author} {\bibfnamefont {B.~P.}\ \bibnamefont
  {Abbott}} \emph {et~al.} (\bibinfo {collaboration} {LIGO Scientific,
  Virgo}),\ }\href@noop {} {\  (\bibinfo {year} {2018})},\ \Eprint
  {http://arxiv.org/abs/1811.12907} {arXiv:1811.12907 [astro-ph.HE]}
  \BibitemShut {NoStop}%
\bibitem [{\citenamefont {Akiyama}\ \emph
  {et~al.}(2019{\natexlab{a}})\citenamefont {Akiyama} \emph
  {et~al.}}]{Akiyama:2019cqa}%
  \BibitemOpen
  \bibfield  {author} {\bibinfo {author} {\bibfnamefont {K.}~\bibnamefont
  {Akiyama}} \emph {et~al.} (\bibinfo {collaboration} {Event Horizon
  Telescope}),\ }\href {\doibase 10.3847/2041-8213/ab0ec7} {\bibfield
  {journal} {\bibinfo  {journal} {Astrophys. J.}\ }\textbf {\bibinfo {volume}
  {875}},\ \bibinfo {pages} {L1} (\bibinfo {year}
  {2019}{\natexlab{a}})}\BibitemShut {NoStop}%
\bibitem [{\citenamefont {Akiyama}\ \emph
  {et~al.}(2019{\natexlab{b}})\citenamefont {Akiyama} \emph
  {et~al.}}]{Akiyama:2019fyp}%
  \BibitemOpen
  \bibfield  {author} {\bibinfo {author} {\bibfnamefont {K.}~\bibnamefont
  {Akiyama}} \emph {et~al.} (\bibinfo {collaboration} {Event Horizon
  Telescope}),\ }\href {\doibase 10.3847/2041-8213/ab0f43} {\bibfield
  {journal} {\bibinfo  {journal} {Astrophys. J.}\ }\textbf {\bibinfo {volume}
  {875}},\ \bibinfo {pages} {L5} (\bibinfo {year}
  {2019}{\natexlab{b}})}\BibitemShut {NoStop}%
\bibitem [{\citenamefont {Akiyama}\ \emph
  {et~al.}(2019{\natexlab{c}})\citenamefont {Akiyama} \emph
  {et~al.}}]{Akiyama:2019eap}%
  \BibitemOpen
  \bibfield  {author} {\bibinfo {author} {\bibfnamefont {K.}~\bibnamefont
  {Akiyama}} \emph {et~al.} (\bibinfo {collaboration} {Event Horizon
  Telescope}),\ }\href {\doibase 10.3847/2041-8213/ab1141} {\bibfield
  {journal} {\bibinfo  {journal} {Astrophys. J.}\ }\textbf {\bibinfo {volume}
  {875}},\ \bibinfo {pages} {L6} (\bibinfo {year}
  {2019}{\natexlab{c}})}\BibitemShut {NoStop}%
\bibitem [{\citenamefont {Kerr}(1963)}]{Kerr:1963ud}%
  \BibitemOpen
  \bibfield  {author} {\bibinfo {author} {\bibfnamefont {R.~P.}\ \bibnamefont
  {Kerr}},\ }\href {\doibase 10.1103/PhysRevLett.11.237} {\bibfield  {journal}
  {\bibinfo  {journal} {Phys.Rev.Lett.}\ }\textbf {\bibinfo {volume} {11}},\
  \bibinfo {pages} {237} (\bibinfo {year} {1963})}\BibitemShut {NoStop}%
\bibitem [{\citenamefont {Berti}\ \emph {et~al.}(2015)\citenamefont {Berti},
  \citenamefont {Barausse}, \citenamefont {Cardoso}, \citenamefont {Gualtieri},
  \citenamefont {Pani} \emph {et~al.}}]{Berti:2015itd}%
  \BibitemOpen
  \bibfield  {author} {\bibinfo {author} {\bibfnamefont {E.}~\bibnamefont
  {Berti}}, \bibinfo {author} {\bibfnamefont {E.}~\bibnamefont {Barausse}},
  \bibinfo {author} {\bibfnamefont {V.}~\bibnamefont {Cardoso}}, \bibinfo
  {author} {\bibfnamefont {L.}~\bibnamefont {Gualtieri}}, \bibinfo {author}
  {\bibfnamefont {P.}~\bibnamefont {Pani}},  \emph {et~al.},\ }\href@noop {} {\
   (\bibinfo {year} {2015})},\ \Eprint {http://arxiv.org/abs/1501.07274}
  {arXiv:1501.07274 [gr-qc]} \BibitemShut {NoStop}%
\bibitem [{\citenamefont {Cardoso}\ and\ \citenamefont
  {Gualtieri}(2016)}]{Cardoso:2016ryw}%
  \BibitemOpen
  \bibfield  {author} {\bibinfo {author} {\bibfnamefont {V.}~\bibnamefont
  {Cardoso}}\ and\ \bibinfo {author} {\bibfnamefont {L.}~\bibnamefont
  {Gualtieri}},\ }\href {\doibase 10.1088/0264-9381/33/17/174001} {\bibfield
  {journal} {\bibinfo  {journal} {Class. Quant. Grav.}\ }\textbf {\bibinfo
  {volume} {33}},\ \bibinfo {pages} {174001} (\bibinfo {year} {2016})},\
  \Eprint {http://arxiv.org/abs/1607.03133} {arXiv:1607.03133 [gr-qc]}
  \BibitemShut {NoStop}%
\bibitem [{\citenamefont {Barack}\ \emph {et~al.}(2018)\citenamefont {Barack}
  \emph {et~al.}}]{Barack:2018yly}%
  \BibitemOpen
  \bibfield  {author} {\bibinfo {author} {\bibfnamefont {L.}~\bibnamefont
  {Barack}} \emph {et~al.},\ }\href@noop {} {\  (\bibinfo {year} {2018})},\
  \Eprint {http://arxiv.org/abs/1806.05195} {arXiv:1806.05195 [gr-qc]}
  \BibitemShut {NoStop}%
\bibitem [{\citenamefont {Degollado}\ \emph {et~al.}(2018)\citenamefont
  {Degollado}, \citenamefont {Herdeiro},\ and\ \citenamefont
  {Radu}}]{Degollado:2018ypf}%
  \BibitemOpen
  \bibfield  {author} {\bibinfo {author} {\bibfnamefont {J.~C.}\ \bibnamefont
  {Degollado}}, \bibinfo {author} {\bibfnamefont {C.~A.~R.}\ \bibnamefont
  {Herdeiro}}, \ and\ \bibinfo {author} {\bibfnamefont {E.}~\bibnamefont
  {Radu}},\ }\href {\doibase 10.1016/j.physletb.2018.04.052} {\bibfield
  {journal} {\bibinfo  {journal} {Phys. Lett.}\ }\textbf {\bibinfo {volume}
  {B781}},\ \bibinfo {pages} {651} (\bibinfo {year} {2018})},\ \Eprint
  {http://arxiv.org/abs/1802.07266} {arXiv:1802.07266 [gr-qc]} \BibitemShut
  {NoStop}%
\bibitem [{\citenamefont {Doneva}\ and\ \citenamefont
  {Yazadjiev}(2018)}]{Doneva:2017bvd}%
  \BibitemOpen
  \bibfield  {author} {\bibinfo {author} {\bibfnamefont {D.~D.}\ \bibnamefont
  {Doneva}}\ and\ \bibinfo {author} {\bibfnamefont {S.~S.}\ \bibnamefont
  {Yazadjiev}},\ }\href {\doibase 10.1103/PhysRevLett.120.131103} {\bibfield
  {journal} {\bibinfo  {journal} {Phys. Rev. Lett.}\ }\textbf {\bibinfo
  {volume} {120}},\ \bibinfo {pages} {131103} (\bibinfo {year} {2018})},\
  \Eprint {http://arxiv.org/abs/1711.01187} {arXiv:1711.01187 [gr-qc]}
  \BibitemShut {NoStop}%
\bibitem [{\citenamefont {Silva}\ \emph {et~al.}(2018)\citenamefont {Silva},
  \citenamefont {Sakstein}, \citenamefont {Gualtieri}, \citenamefont
  {Sotiriou},\ and\ \citenamefont {Berti}}]{Silva:2017uqg}%
  \BibitemOpen
  \bibfield  {author} {\bibinfo {author} {\bibfnamefont {H.~O.}\ \bibnamefont
  {Silva}}, \bibinfo {author} {\bibfnamefont {J.}~\bibnamefont {Sakstein}},
  \bibinfo {author} {\bibfnamefont {L.}~\bibnamefont {Gualtieri}}, \bibinfo
  {author} {\bibfnamefont {T.~P.}\ \bibnamefont {Sotiriou}}, \ and\ \bibinfo
  {author} {\bibfnamefont {E.}~\bibnamefont {Berti}},\ }\href {\doibase
  10.1103/PhysRevLett.120.131104} {\bibfield  {journal} {\bibinfo  {journal}
  {Phys. Rev. Lett.}\ }\textbf {\bibinfo {volume} {120}},\ \bibinfo {pages}
  {131104} (\bibinfo {year} {2018})},\ \Eprint
  {http://arxiv.org/abs/1711.02080} {arXiv:1711.02080 [gr-qc]} \BibitemShut
  {NoStop}%
\bibitem [{\citenamefont {Antoniou}\ \emph
  {et~al.}(2018{\natexlab{a}})\citenamefont {Antoniou}, \citenamefont
  {Bakopoulos},\ and\ \citenamefont {Kanti}}]{Antoniou:2017acq}%
  \BibitemOpen
  \bibfield  {author} {\bibinfo {author} {\bibfnamefont {G.}~\bibnamefont
  {Antoniou}}, \bibinfo {author} {\bibfnamefont {A.}~\bibnamefont
  {Bakopoulos}}, \ and\ \bibinfo {author} {\bibfnamefont {P.}~\bibnamefont
  {Kanti}},\ }\href {\doibase 10.1103/PhysRevLett.120.131102} {\bibfield
  {journal} {\bibinfo  {journal} {Phys. Rev. Lett.}\ }\textbf {\bibinfo
  {volume} {120}},\ \bibinfo {pages} {131102} (\bibinfo {year}
  {2018}{\natexlab{a}})},\ \Eprint {http://arxiv.org/abs/1711.03390}
  {arXiv:1711.03390 [hep-th]} \BibitemShut {NoStop}%
\bibitem [{\citenamefont {Damour}\ and\ \citenamefont
  {Esposito-Farese}(1993)}]{Damour:1993hw}%
  \BibitemOpen
  \bibfield  {author} {\bibinfo {author} {\bibfnamefont {T.}~\bibnamefont
  {Damour}}\ and\ \bibinfo {author} {\bibfnamefont {G.}~\bibnamefont
  {Esposito-Farese}},\ }\href {\doibase 10.1103/PhysRevLett.70.2220} {\bibfield
   {journal} {\bibinfo  {journal} {Phys. Rev. Lett.}\ }\textbf {\bibinfo
  {volume} {70}},\ \bibinfo {pages} {2220} (\bibinfo {year}
  {1993})}\BibitemShut {NoStop}%
\bibitem [{\citenamefont {Ostrogradsky}(1850)}]{Ostrogradsky:1850fid}%
  \BibitemOpen
  \bibfield  {author} {\bibinfo {author} {\bibfnamefont {M.}~\bibnamefont
  {Ostrogradsky}},\ }\href@noop {} {\bibfield  {journal} {\bibinfo  {journal}
  {Mem. Acad. St. Petersbourg}\ }\textbf {\bibinfo {volume} {6}},\ \bibinfo
  {pages} {385} (\bibinfo {year} {1850})}\BibitemShut {NoStop}%
\bibitem [{\citenamefont {Zwiebach}(1985)}]{Zwiebach:1985uq}%
  \BibitemOpen
  \bibfield  {author} {\bibinfo {author} {\bibfnamefont {B.}~\bibnamefont
  {Zwiebach}},\ }\href {\doibase 10.1016/0370-2693(85)91616-8} {\bibfield
  {journal} {\bibinfo  {journal} {Phys. Lett.}\ }\textbf {\bibinfo {volume}
  {156B}},\ \bibinfo {pages} {315} (\bibinfo {year} {1985})}\BibitemShut
  {NoStop}%
\bibitem [{\citenamefont {Antoniou}\ \emph
  {et~al.}(2018{\natexlab{b}})\citenamefont {Antoniou}, \citenamefont
  {Bakopoulos},\ and\ \citenamefont {Kanti}}]{Antoniou:2017hxj}%
  \BibitemOpen
  \bibfield  {author} {\bibinfo {author} {\bibfnamefont {G.}~\bibnamefont
  {Antoniou}}, \bibinfo {author} {\bibfnamefont {A.}~\bibnamefont
  {Bakopoulos}}, \ and\ \bibinfo {author} {\bibfnamefont {P.}~\bibnamefont
  {Kanti}},\ }\href {\doibase 10.1103/PhysRevD.97.084037} {\bibfield  {journal}
  {\bibinfo  {journal} {Phys. Rev.}\ }\textbf {\bibinfo {volume} {D97}},\
  \bibinfo {pages} {084037} (\bibinfo {year} {2018}{\natexlab{b}})},\ \Eprint
  {http://arxiv.org/abs/1711.07431} {arXiv:1711.07431 [hep-th]} \BibitemShut
  {NoStop}%
\bibitem [{\citenamefont {Minamitsuji}\ and\ \citenamefont
  {Ikeda}(2019)}]{Minamitsuji:2018xde}%
  \BibitemOpen
  \bibfield  {author} {\bibinfo {author} {\bibfnamefont {M.}~\bibnamefont
  {Minamitsuji}}\ and\ \bibinfo {author} {\bibfnamefont {T.}~\bibnamefont
  {Ikeda}},\ }\href {\doibase 10.1103/PhysRevD.99.044017} {\bibfield  {journal}
  {\bibinfo  {journal} {Phys. Rev.}\ }\textbf {\bibinfo {volume} {D99}},\
  \bibinfo {pages} {044017} (\bibinfo {year} {2019})},\ \Eprint
  {http://arxiv.org/abs/1812.03551} {arXiv:1812.03551 [gr-qc]} \BibitemShut
  {NoStop}%
\bibitem [{\citenamefont {Bakopoulos}\ \emph {et~al.}(2019)\citenamefont
  {Bakopoulos}, \citenamefont {Antoniou},\ and\ \citenamefont
  {Kanti}}]{Bakopoulos:2018nui}%
  \BibitemOpen
  \bibfield  {author} {\bibinfo {author} {\bibfnamefont {A.}~\bibnamefont
  {Bakopoulos}}, \bibinfo {author} {\bibfnamefont {G.}~\bibnamefont
  {Antoniou}}, \ and\ \bibinfo {author} {\bibfnamefont {P.}~\bibnamefont
  {Kanti}},\ }\href {\doibase 10.1103/PhysRevD.99.064003} {\bibfield  {journal}
  {\bibinfo  {journal} {Phys. Rev.}\ }\textbf {\bibinfo {volume} {D99}},\
  \bibinfo {pages} {064003} (\bibinfo {year} {2019})},\ \Eprint
  {http://arxiv.org/abs/1812.06941} {arXiv:1812.06941 [hep-th]} \BibitemShut
  {NoStop}%
\bibitem [{\citenamefont {Motohashi}\ and\ \citenamefont
  {Mukohyama}(2019)}]{Motohashi:2018mql}%
  \BibitemOpen
  \bibfield  {author} {\bibinfo {author} {\bibfnamefont {H.}~\bibnamefont
  {Motohashi}}\ and\ \bibinfo {author} {\bibfnamefont {S.}~\bibnamefont
  {Mukohyama}},\ }\href {\doibase 10.1103/PhysRevD.99.044030} {\bibfield
  {journal} {\bibinfo  {journal} {Phys. Rev.}\ }\textbf {\bibinfo {volume}
  {D99}},\ \bibinfo {pages} {044030} (\bibinfo {year} {2019})},\ \Eprint
  {http://arxiv.org/abs/1810.12691} {arXiv:1810.12691 [gr-qc]} \BibitemShut
  {NoStop}%
\bibitem [{\citenamefont {Brihaye}\ and\ \citenamefont
  {Ducobu}(2018)}]{Brihaye:2018grv}%
  \BibitemOpen
  \bibfield  {author} {\bibinfo {author} {\bibfnamefont {Y.}~\bibnamefont
  {Brihaye}}\ and\ \bibinfo {author} {\bibfnamefont {L.}~\bibnamefont
  {Ducobu}},\ }\href@noop {} {\  (\bibinfo {year} {2018})},\ \Eprint
  {http://arxiv.org/abs/1812.07438} {arXiv:1812.07438 [gr-qc]} \BibitemShut
  {NoStop}%
\bibitem [{\citenamefont {Macedo}\ \emph {et~al.}(2019)\citenamefont {Macedo},
  \citenamefont {Sakstein}, \citenamefont {Berti}, \citenamefont {Gualtieri},
  \citenamefont {Silva},\ and\ \citenamefont {Sotiriou}}]{Macedo:2019sem}%
  \BibitemOpen
  \bibfield  {author} {\bibinfo {author} {\bibfnamefont {C.~F.~B.}\
  \bibnamefont {Macedo}}, \bibinfo {author} {\bibfnamefont {J.}~\bibnamefont
  {Sakstein}}, \bibinfo {author} {\bibfnamefont {E.}~\bibnamefont {Berti}},
  \bibinfo {author} {\bibfnamefont {L.}~\bibnamefont {Gualtieri}}, \bibinfo
  {author} {\bibfnamefont {H.~O.}\ \bibnamefont {Silva}}, \ and\ \bibinfo
  {author} {\bibfnamefont {T.~P.}\ \bibnamefont {Sotiriou}},\ }\href@noop {} {\
   (\bibinfo {year} {2019})},\ \Eprint {http://arxiv.org/abs/1903.06784}
  {arXiv:1903.06784 [gr-qc]} \BibitemShut {NoStop}%
\bibitem [{\citenamefont {Doneva}\ \emph {et~al.}(2019)\citenamefont {Doneva},
  \citenamefont {Staykov},\ and\ \citenamefont {Yazadjiev}}]{Doneva:2019vuh}%
  \BibitemOpen
  \bibfield  {author} {\bibinfo {author} {\bibfnamefont {D.~D.}\ \bibnamefont
  {Doneva}}, \bibinfo {author} {\bibfnamefont {K.~V.}\ \bibnamefont {Staykov}},
  \ and\ \bibinfo {author} {\bibfnamefont {S.~S.}\ \bibnamefont {Yazadjiev}},\
  }\href@noop {} {\  (\bibinfo {year} {2019})},\ \Eprint
  {http://arxiv.org/abs/1903.08119} {arXiv:1903.08119 [gr-qc]} \BibitemShut
  {NoStop}%
\bibitem [{\citenamefont {Myung}\ and\ \citenamefont
  {Zou}(2019{\natexlab{a}})}]{Myung:2019wvb}%
  \BibitemOpen
  \bibfield  {author} {\bibinfo {author} {\bibfnamefont {Y.~S.}\ \bibnamefont
  {Myung}}\ and\ \bibinfo {author} {\bibfnamefont {D.-C.}\ \bibnamefont
  {Zou}},\ }\href@noop {} {\  (\bibinfo {year} {2019}{\natexlab{a}})},\ \Eprint
  {http://arxiv.org/abs/1903.08312} {arXiv:1903.08312 [gr-qc]} \BibitemShut
  {NoStop}%
\bibitem [{\citenamefont {Brihaye}\ and\ \citenamefont
  {Hartmann}(2019)}]{Brihaye:2019kvj}%
  \BibitemOpen
  \bibfield  {author} {\bibinfo {author} {\bibfnamefont {Y.}~\bibnamefont
  {Brihaye}}\ and\ \bibinfo {author} {\bibfnamefont {B.}~\bibnamefont
  {Hartmann}},\ }\href {\doibase 10.1016/j.physletb.2019.03.043} {\  (\bibinfo
  {year} {2019}),\ 10.1016/j.physletb.2019.03.043},\ \Eprint
  {http://arxiv.org/abs/1902.05760} {arXiv:1902.05760 [gr-qc]} \BibitemShut
  {NoStop}%
\bibitem [{\citenamefont {Blázquez-Salcedo}\ \emph {et~al.}(2018)\citenamefont
  {Blázquez-Salcedo}, \citenamefont {Doneva}, \citenamefont {Kunz},\ and\
  \citenamefont {Yazadjiev}}]{Blazquez-Salcedo:2018jnn}%
  \BibitemOpen
  \bibfield  {author} {\bibinfo {author} {\bibfnamefont {J.~L.}\ \bibnamefont
  {Blázquez-Salcedo}}, \bibinfo {author} {\bibfnamefont {D.~D.}\ \bibnamefont
  {Doneva}}, \bibinfo {author} {\bibfnamefont {J.}~\bibnamefont {Kunz}}, \ and\
  \bibinfo {author} {\bibfnamefont {S.~S.}\ \bibnamefont {Yazadjiev}},\ }\href
  {\doibase 10.1103/PhysRevD.98.084011} {\bibfield  {journal} {\bibinfo
  {journal} {Phys. Rev.}\ }\textbf {\bibinfo {volume} {D98}},\ \bibinfo {pages}
  {084011} (\bibinfo {year} {2018})},\ \Eprint
  {http://arxiv.org/abs/1805.05755} {arXiv:1805.05755 [gr-qc]} \BibitemShut
  {NoStop}%
\bibitem [{\citenamefont {Silva}\ \emph {et~al.}(2019)\citenamefont {Silva},
  \citenamefont {Macedo}, \citenamefont {Sotiriou}, \citenamefont {Gualtieri},
  \citenamefont {Sakstein},\ and\ \citenamefont {Berti}}]{Silva:2018qhn}%
  \BibitemOpen
  \bibfield  {author} {\bibinfo {author} {\bibfnamefont {H.~O.}\ \bibnamefont
  {Silva}}, \bibinfo {author} {\bibfnamefont {C.~F.~B.}\ \bibnamefont
  {Macedo}}, \bibinfo {author} {\bibfnamefont {T.~P.}\ \bibnamefont
  {Sotiriou}}, \bibinfo {author} {\bibfnamefont {L.}~\bibnamefont {Gualtieri}},
  \bibinfo {author} {\bibfnamefont {J.}~\bibnamefont {Sakstein}}, \ and\
  \bibinfo {author} {\bibfnamefont {E.}~\bibnamefont {Berti}},\ }\href
  {\doibase 10.1103/PhysRevD.99.064011} {\bibfield  {journal} {\bibinfo
  {journal} {Phys. Rev.}\ }\textbf {\bibinfo {volume} {D99}},\ \bibinfo {pages}
  {064011} (\bibinfo {year} {2019})},\ \Eprint
  {http://arxiv.org/abs/1812.05590} {arXiv:1812.05590 [gr-qc]} \BibitemShut
  {NoStop}%
\bibitem [{\citenamefont {Sotiriou}\ and\ \citenamefont
  {Faraoni}(2012)}]{Sotiriou:2011dz}%
  \BibitemOpen
  \bibfield  {author} {\bibinfo {author} {\bibfnamefont {T.~P.}\ \bibnamefont
  {Sotiriou}}\ and\ \bibinfo {author} {\bibfnamefont {V.}~\bibnamefont
  {Faraoni}},\ }\href {\doibase 10.1103/PhysRevLett.108.081103} {\bibfield
  {journal} {\bibinfo  {journal} {Phys. Rev. Lett.}\ }\textbf {\bibinfo
  {volume} {108}},\ \bibinfo {pages} {081103} (\bibinfo {year} {2012})},\
  \Eprint {http://arxiv.org/abs/1109.6324} {arXiv:1109.6324 [gr-qc]}
  \BibitemShut {NoStop}%
\bibitem [{\citenamefont {Herdeiro}\ and\ \citenamefont
  {Radu}(2015{\natexlab{a}})}]{Herdeiro:2015waa}%
  \BibitemOpen
  \bibfield  {author} {\bibinfo {author} {\bibfnamefont {C.~A.~R.}\
  \bibnamefont {Herdeiro}}\ and\ \bibinfo {author} {\bibfnamefont
  {E.}~\bibnamefont {Radu}},\ }\bibfield  {booktitle} {\emph {\bibinfo
  {booktitle} {{Proceedings, 7th Black Holes Workshop 2014: Aveiro, Portugal,
  December 18-19, 2014}}},\ }\href {\doibase 10.1142/S0218271815420146}
  {\bibfield  {journal} {\bibinfo  {journal} {Int. J. Mod. Phys.}\ }\textbf
  {\bibinfo {volume} {D24}},\ \bibinfo {pages} {1542014} (\bibinfo {year}
  {2015}{\natexlab{a}})},\ \Eprint {http://arxiv.org/abs/1504.08209}
  {arXiv:1504.08209 [gr-qc]} \BibitemShut {NoStop}%
\bibitem [{\citenamefont {Dirac}(1938)}]{Dirac:1938mt}%
  \BibitemOpen
  \bibfield  {author} {\bibinfo {author} {\bibfnamefont {P.~A.~M.}\
  \bibnamefont {Dirac}},\ }\href {\doibase 10.1098/rspa.1938.0053} {\bibfield
  {journal} {\bibinfo  {journal} {Proc. Roy. Soc. Lond.}\ }\textbf {\bibinfo
  {volume} {A165}},\ \bibinfo {pages} {199} (\bibinfo {year}
  {1938})}\BibitemShut {NoStop}%
\bibitem [{\citenamefont {Brans}\ and\ \citenamefont
  {Dicke}(1961)}]{Brans:1961sx}%
  \BibitemOpen
  \bibfield  {author} {\bibinfo {author} {\bibfnamefont {C.}~\bibnamefont
  {Brans}}\ and\ \bibinfo {author} {\bibfnamefont {R.~H.}\ \bibnamefont
  {Dicke}},\ }\href {\doibase 10.1103/PhysRev.124.925} {\bibfield  {journal}
  {\bibinfo  {journal} {Phys. Rev.}\ }\textbf {\bibinfo {volume} {124}},\
  \bibinfo {pages} {925} (\bibinfo {year} {1961})}\BibitemShut {NoStop}%
\bibitem [{\citenamefont {Herdeiro}\ \emph {et~al.}(2018)\citenamefont
  {Herdeiro}, \citenamefont {Radu}, \citenamefont {Sanchis-Gual},\ and\
  \citenamefont {Font}}]{Herdeiro:2018wub}%
  \BibitemOpen
  \bibfield  {author} {\bibinfo {author} {\bibfnamefont {C.~A.~R.}\
  \bibnamefont {Herdeiro}}, \bibinfo {author} {\bibfnamefont {E.}~\bibnamefont
  {Radu}}, \bibinfo {author} {\bibfnamefont {N.}~\bibnamefont {Sanchis-Gual}},
  \ and\ \bibinfo {author} {\bibfnamefont {J.~A.}\ \bibnamefont {Font}},\
  }\href {\doibase 10.1103/PhysRevLett.121.101102} {\bibfield  {journal}
  {\bibinfo  {journal} {Phys. Rev. Lett.}\ }\textbf {\bibinfo {volume} {121}},\
  \bibinfo {pages} {101102} (\bibinfo {year} {2018})},\ \Eprint
  {http://arxiv.org/abs/1806.05190} {arXiv:1806.05190 [gr-qc]} \BibitemShut
  {NoStop}%
\bibitem [{\citenamefont {Myung}\ and\ \citenamefont
  {Zou}(2019{\natexlab{b}})}]{Myung:2018jvi}%
  \BibitemOpen
  \bibfield  {author} {\bibinfo {author} {\bibfnamefont {Y.~S.}\ \bibnamefont
  {Myung}}\ and\ \bibinfo {author} {\bibfnamefont {D.-C.}\ \bibnamefont
  {Zou}},\ }\href {\doibase 10.1016/j.physletb.2019.01.046} {\bibfield
  {journal} {\bibinfo  {journal} {Phys. Lett.}\ }\textbf {\bibinfo {volume}
  {B790}},\ \bibinfo {pages} {400} (\bibinfo {year} {2019}{\natexlab{b}})},\
  \Eprint {http://arxiv.org/abs/1812.03604} {arXiv:1812.03604 [gr-qc]}
  \BibitemShut {NoStop}%
\bibitem [{\citenamefont {Fernandes}\ \emph {et~al.}(2019)\citenamefont
  {Fernandes}, \citenamefont {Herdeiro}, \citenamefont {Pombo}, \citenamefont
  {Radu},\ and\ \citenamefont {Sanchis-Gual}}]{Fernandes:2019rez}%
  \BibitemOpen
  \bibfield  {author} {\bibinfo {author} {\bibfnamefont {P.~G.~S.}\
  \bibnamefont {Fernandes}}, \bibinfo {author} {\bibfnamefont {C.~A.~R.}\
  \bibnamefont {Herdeiro}}, \bibinfo {author} {\bibfnamefont {A.~M.}\
  \bibnamefont {Pombo}}, \bibinfo {author} {\bibfnamefont {E.}~\bibnamefont
  {Radu}}, \ and\ \bibinfo {author} {\bibfnamefont {N.}~\bibnamefont
  {Sanchis-Gual}},\ }\href@noop {} {\  (\bibinfo {year} {2019})},\ \Eprint
  {http://arxiv.org/abs/1902.05079} {arXiv:1902.05079 [gr-qc]} \BibitemShut
  {NoStop}%
\bibitem [{\citenamefont {Herdeiro}\ and\ \citenamefont
  {Radu}(2015{\natexlab{b}})}]{Herdeiro:2015gia}%
  \BibitemOpen
  \bibfield  {author} {\bibinfo {author} {\bibfnamefont {C.}~\bibnamefont
  {Herdeiro}}\ and\ \bibinfo {author} {\bibfnamefont {E.}~\bibnamefont
  {Radu}},\ }\href {\doibase 10.1088/0264-9381/32/14/144001} {\bibfield
  {journal} {\bibinfo  {journal} {Class.Quant.Grav.}\ }\textbf {\bibinfo
  {volume} {32}},\ \bibinfo {pages} {144001} (\bibinfo {year}
  {2015}{\natexlab{b}})},\ \Eprint {http://arxiv.org/abs/1501.04319}
  {arXiv:1501.04319 [gr-qc]} \BibitemShut {NoStop}%
\bibitem [{\citenamefont {Kanti}\ \emph {et~al.}(1996)\citenamefont {Kanti},
  \citenamefont {Mavromatos}, \citenamefont {Rizos}, \citenamefont {Tamvakis},\
  and\ \citenamefont {Winstanley}}]{Kanti:1995vq}%
  \BibitemOpen
  \bibfield  {author} {\bibinfo {author} {\bibfnamefont {P.}~\bibnamefont
  {Kanti}}, \bibinfo {author} {\bibfnamefont {N.~E.}\ \bibnamefont
  {Mavromatos}}, \bibinfo {author} {\bibfnamefont {J.}~\bibnamefont {Rizos}},
  \bibinfo {author} {\bibfnamefont {K.}~\bibnamefont {Tamvakis}}, \ and\
  \bibinfo {author} {\bibfnamefont {E.}~\bibnamefont {Winstanley}},\ }\href
  {\doibase 10.1103/PhysRevD.54.5049} {\bibfield  {journal} {\bibinfo
  {journal} {Phys. Rev.}\ }\textbf {\bibinfo {volume} {D54}},\ \bibinfo {pages}
  {5049} (\bibinfo {year} {1996})},\ \Eprint
  {http://arxiv.org/abs/hep-th/9511071} {arXiv:hep-th/9511071 [hep-th]}
  \BibitemShut {NoStop}%
\bibitem [{\citenamefont {Kleihaus}\ \emph {et~al.}(2016)\citenamefont
  {Kleihaus}, \citenamefont {Kunz}, \citenamefont {Mojica},\ and\ \citenamefont
  {Radu}}]{Kleihaus:2015aje}%
  \BibitemOpen
  \bibfield  {author} {\bibinfo {author} {\bibfnamefont {B.}~\bibnamefont
  {Kleihaus}}, \bibinfo {author} {\bibfnamefont {J.}~\bibnamefont {Kunz}},
  \bibinfo {author} {\bibfnamefont {S.}~\bibnamefont {Mojica}}, \ and\ \bibinfo
  {author} {\bibfnamefont {E.}~\bibnamefont {Radu}},\ }\href {\doibase
  10.1103/PhysRevD.93.044047} {\bibfield  {journal} {\bibinfo  {journal} {Phys.
  Rev.}\ }\textbf {\bibinfo {volume} {D93}},\ \bibinfo {pages} {044047}
  (\bibinfo {year} {2016})},\ \Eprint {http://arxiv.org/abs/1511.05513}
  {arXiv:1511.05513 [gr-qc]} \BibitemShut {NoStop}%
\bibitem [{Note1()}]{Note1}%
  \BibitemOpen
  \bibinfo {note} {Technically this limit can be understood from the
  perturbative expansion of the solutions near the horizon, wherein the
  radicand of a square root becomes negative.}\BibitemShut {Stop}%
\bibitem [{\citenamefont {Komar}(1959)}]{Komar:1958wp}%
  \BibitemOpen
  \bibfield  {author} {\bibinfo {author} {\bibfnamefont {A.}~\bibnamefont
  {Komar}},\ }\href {\doibase 10.1103/PhysRev.113.934} {\bibfield  {journal}
  {\bibinfo  {journal} {Phys. Rev.}\ }\textbf {\bibinfo {volume} {113}},\
  \bibinfo {pages} {934} (\bibinfo {year} {1959})}\BibitemShut {NoStop}%
\bibitem [{\citenamefont {Wald}(1993)}]{Wald:1993nt}%
  \BibitemOpen
  \bibfield  {author} {\bibinfo {author} {\bibfnamefont {R.~M.}\ \bibnamefont
  {Wald}},\ }\href {\doibase 10.1103/PhysRevD.48.R3427} {\bibfield  {journal}
  {\bibinfo  {journal} {Phys. Rev.}\ }\textbf {\bibinfo {volume} {D48}},\
  \bibinfo {pages} {R3427} (\bibinfo {year} {1993})},\ \Eprint
  {http://arxiv.org/abs/gr-qc/9307038} {arXiv:gr-qc/9307038 [gr-qc]}
  \BibitemShut {NoStop}%
\bibitem [{\citenamefont {Smarr}(1973)}]{Smarr:1972kt}%
  \BibitemOpen
  \bibfield  {author} {\bibinfo {author} {\bibfnamefont {L.}~\bibnamefont
  {Smarr}},\ }\href {\doibase 10.1103/PhysRevLett.30.521,
  10.1103/PhysRevLett.30.71} {\bibfield  {journal} {\bibinfo  {journal} {Phys.
  Rev. Lett.}\ }\textbf {\bibinfo {volume} {30}},\ \bibinfo {pages} {71}
  (\bibinfo {year} {1973})},\ \bibinfo {note} {[Erratum: Phys. Rev.
  Lett.30,521(1973)]}\BibitemShut {NoStop}%
\bibitem [{\citenamefont {Perlick}(2004)}]{Perlick:2004tq}%
  \BibitemOpen
  \bibfield  {author} {\bibinfo {author} {\bibfnamefont {V.}~\bibnamefont
  {Perlick}},\ }\href@noop {} {\bibfield  {journal} {\bibinfo  {journal}
  {Living Rev. Rel.}\ }\textbf {\bibinfo {volume} {7}},\ \bibinfo {pages} {9}
  (\bibinfo {year} {2004})}\BibitemShut {NoStop}%
\bibitem [{\citenamefont {Cunha}\ and\ \citenamefont
  {Herdeiro}(2018)}]{Cunha:2018acu}%
  \BibitemOpen
  \bibfield  {author} {\bibinfo {author} {\bibfnamefont {P.~V.~P.}\
  \bibnamefont {Cunha}}\ and\ \bibinfo {author} {\bibfnamefont {C.~A.~R.}\
  \bibnamefont {Herdeiro}},\ }\href {\doibase 10.1007/s10714-018-2361-9}
  {\bibfield  {journal} {\bibinfo  {journal} {Gen. Rel. Grav.}\ }\textbf
  {\bibinfo {volume} {50}},\ \bibinfo {pages} {42} (\bibinfo {year} {2018})},\
  \Eprint {http://arxiv.org/abs/1801.00860} {arXiv:1801.00860 [gr-qc]}
  \BibitemShut {NoStop}%
\bibitem [{\citenamefont {Cunha}\ \emph
  {et~al.}(2017{\natexlab{a}})\citenamefont {Cunha}, \citenamefont {Herdeiro},
  \citenamefont {Kleihaus}, \citenamefont {Kunz},\ and\ \citenamefont
  {Radu}}]{Cunha:2016wzk}%
  \BibitemOpen
  \bibfield  {author} {\bibinfo {author} {\bibfnamefont {P.~V.~P.}\
  \bibnamefont {Cunha}}, \bibinfo {author} {\bibfnamefont {C.~A.~R.}\
  \bibnamefont {Herdeiro}}, \bibinfo {author} {\bibfnamefont {B.}~\bibnamefont
  {Kleihaus}}, \bibinfo {author} {\bibfnamefont {J.}~\bibnamefont {Kunz}}, \
  and\ \bibinfo {author} {\bibfnamefont {E.}~\bibnamefont {Radu}},\ }\href
  {\doibase 10.1016/j.physletb.2017.03.020} {\bibfield  {journal} {\bibinfo
  {journal} {Phys. Lett.}\ }\textbf {\bibinfo {volume} {B768}},\ \bibinfo
  {pages} {373} (\bibinfo {year} {2017}{\natexlab{a}})},\ \Eprint
  {http://arxiv.org/abs/1701.00079} {arXiv:1701.00079 [gr-qc]} \BibitemShut
  {NoStop}%
\bibitem [{\citenamefont {Falcke}\ \emph {et~al.}(2000)\citenamefont {Falcke},
  \citenamefont {Melia},\ and\ \citenamefont {Agol}}]{Falcke:1999pj}%
  \BibitemOpen
  \bibfield  {author} {\bibinfo {author} {\bibfnamefont {H.}~\bibnamefont
  {Falcke}}, \bibinfo {author} {\bibfnamefont {F.}~\bibnamefont {Melia}}, \
  and\ \bibinfo {author} {\bibfnamefont {E.}~\bibnamefont {Agol}},\ }\href
  {\doibase 10.1086/312423} {\bibfield  {journal} {\bibinfo  {journal}
  {Astrophys.J.}\ }\textbf {\bibinfo {volume} {528}},\ \bibinfo {pages} {L13}
  (\bibinfo {year} {2000})},\ \Eprint {http://arxiv.org/abs/astro-ph/9912263}
  {arXiv:astro-ph/9912263 [astro-ph]} \BibitemShut {NoStop}%
\bibitem [{\citenamefont {Johannsen}\ \emph {et~al.}(2016)\citenamefont
  {Johannsen}, \citenamefont {Wang}, \citenamefont {Broderick}, \citenamefont
  {Doeleman}, \citenamefont {Fish}, \citenamefont {Loeb},\ and\ \citenamefont
  {Psaltis}}]{Johannsen:2016vqy}%
  \BibitemOpen
  \bibfield  {author} {\bibinfo {author} {\bibfnamefont {T.}~\bibnamefont
  {Johannsen}}, \bibinfo {author} {\bibfnamefont {C.}~\bibnamefont {Wang}},
  \bibinfo {author} {\bibfnamefont {A.~E.}\ \bibnamefont {Broderick}}, \bibinfo
  {author} {\bibfnamefont {S.~S.}\ \bibnamefont {Doeleman}}, \bibinfo {author}
  {\bibfnamefont {V.~L.}\ \bibnamefont {Fish}}, \bibinfo {author}
  {\bibfnamefont {A.}~\bibnamefont {Loeb}}, \ and\ \bibinfo {author}
  {\bibfnamefont {D.}~\bibnamefont {Psaltis}},\ }\href {\doibase
  10.1103/PhysRevLett.117.091101} {\bibfield  {journal} {\bibinfo  {journal}
  {Phys. Rev. Lett.}\ }\textbf {\bibinfo {volume} {117}},\ \bibinfo {pages}
  {091101} (\bibinfo {year} {2016})},\ \Eprint
  {http://arxiv.org/abs/1608.03593} {arXiv:1608.03593 [astro-ph.HE]}
  \BibitemShut {NoStop}%
\bibitem [{\citenamefont {Cunha}\ \emph
  {et~al.}(2017{\natexlab{b}})\citenamefont {Cunha}, \citenamefont {Herdeiro},\
  and\ \citenamefont {Radu}}]{Cunha:2017eoe}%
  \BibitemOpen
  \bibfield  {author} {\bibinfo {author} {\bibfnamefont {P.~V.~P.}\
  \bibnamefont {Cunha}}, \bibinfo {author} {\bibfnamefont {C.~A.~R.}\
  \bibnamefont {Herdeiro}}, \ and\ \bibinfo {author} {\bibfnamefont
  {E.}~\bibnamefont {Radu}},\ }\href {\doibase 10.1103/PhysRevD.96.024039}
  {\bibfield  {journal} {\bibinfo  {journal} {Phys. Rev.}\ }\textbf {\bibinfo
  {volume} {D96}},\ \bibinfo {pages} {024039} (\bibinfo {year}
  {2017}{\natexlab{b}})},\ \Eprint {http://arxiv.org/abs/1705.05461}
  {arXiv:1705.05461 [gr-qc]} \BibitemShut {NoStop}%
\bibitem [{\citenamefont {Cunha}\ \emph {et~al.}(2016)\citenamefont {Cunha},
  \citenamefont {Herdeiro}, \citenamefont {Radu},\ and\ \citenamefont
  {Runarsson}}]{Cunha:2016bpi}%
  \BibitemOpen
  \bibfield  {author} {\bibinfo {author} {\bibfnamefont {P.~V.~P.}\
  \bibnamefont {Cunha}}, \bibinfo {author} {\bibfnamefont {C.~A.~R.}\
  \bibnamefont {Herdeiro}}, \bibinfo {author} {\bibfnamefont {E.}~\bibnamefont
  {Radu}}, \ and\ \bibinfo {author} {\bibfnamefont {H.~F.}\ \bibnamefont
  {Runarsson}},\ }\bibfield  {booktitle} {\emph {\bibinfo {booktitle}
  {{Proceedings, 3rd Amazonian Symposium on Physics: Belem, Brazil, September
  28-October 2, 2015}}},\ }\href {\doibase 10.1142/S0218271816410212}
  {\bibfield  {journal} {\bibinfo  {journal} {Int. J. Mod. Phys.}\ }\textbf
  {\bibinfo {volume} {D25}},\ \bibinfo {pages} {1641021} (\bibinfo {year}
  {2016})},\ \Eprint {http://arxiv.org/abs/1605.08293} {arXiv:1605.08293
  [gr-qc]} \BibitemShut {NoStop}%
\bibitem [{\citenamefont {NASA/Hubble}()}]{image}%
  \BibitemOpen
  \bibfield  {author} {\bibinfo {author} {\bibnamefont {NASA/Hubble}},\
  }\href@noop {} {\emph {\bibinfo {title} {Background image used for lensing,
  http://hubblesite.org/image/3905/printshop}}}\BibitemShut {NoStop}%
\bibitem [{Note2()}]{Note2}%
  \BibitemOpen
  \bibinfo {note} {The corresponding Schwarzschild shadow \protect \textit
  {diameter} is $6\protect \sqrt {3}\protect \tmspace +\thinmuskip
  {.1667em}\theta _g\simeq 39$ $\mu $as. For a Kerr BH this value varies,
  increasing $j$, by $\lesssim $ 2\%, due to the spin inclination angle $\sim $
  17$^o$. There is an offset between such light ring diameter and the \protect
  \textit {measured} emission ring of $42\pm 3$ $\mu $as; $\theta _g$ is
  inferred from the latter via GRMHD models~\cite {Akiyama:2019eap} which
  predict a peak emission slightly outside the light ring.}\BibitemShut {Stop}%
\bibitem [{\citenamefont {Gebhardt}\ \emph {et~al.}(2011)\citenamefont
  {Gebhardt}, \citenamefont {Adams}, \citenamefont {Richstone}, \citenamefont
  {Lauer}, \citenamefont {Faber}, \citenamefont {Gultekin}, \citenamefont
  {Murphy},\ and\ \citenamefont {Tremaine}}]{Gebhardt:2011yw}%
  \BibitemOpen
  \bibfield  {author} {\bibinfo {author} {\bibfnamefont {K.}~\bibnamefont
  {Gebhardt}}, \bibinfo {author} {\bibfnamefont {J.}~\bibnamefont {Adams}},
  \bibinfo {author} {\bibfnamefont {D.}~\bibnamefont {Richstone}}, \bibinfo
  {author} {\bibfnamefont {T.~R.}\ \bibnamefont {Lauer}}, \bibinfo {author}
  {\bibfnamefont {S.~M.}\ \bibnamefont {Faber}}, \bibinfo {author}
  {\bibfnamefont {K.}~\bibnamefont {Gultekin}}, \bibinfo {author}
  {\bibfnamefont {J.}~\bibnamefont {Murphy}}, \ and\ \bibinfo {author}
  {\bibfnamefont {S.}~\bibnamefont {Tremaine}},\ }\href {\doibase
  10.1088/0004-637X/729/2/119} {\bibfield  {journal} {\bibinfo  {journal}
  {Astrophys. J.}\ }\textbf {\bibinfo {volume} {729}},\ \bibinfo {pages} {119}
  (\bibinfo {year} {2011})},\ \Eprint {http://arxiv.org/abs/1101.1954}
  {arXiv:1101.1954 [astro-ph.CO]} \BibitemShut {NoStop}%
\bibitem [{\citenamefont {Nokhrina}\ \emph {et~al.}(2019)\citenamefont
  {Nokhrina}, \citenamefont {Gurvits}, \citenamefont {Beskin}, \citenamefont
  {Nakamura}, \citenamefont {Asada},\ and\ \citenamefont
  {Hada}}]{Nokhrina:2019sxv}%
  \BibitemOpen
  \bibfield  {author} {\bibinfo {author} {\bibfnamefont {E.~E.}\ \bibnamefont
  {Nokhrina}}, \bibinfo {author} {\bibfnamefont {L.~I.}\ \bibnamefont
  {Gurvits}}, \bibinfo {author} {\bibfnamefont {V.~S.}\ \bibnamefont {Beskin}},
  \bibinfo {author} {\bibfnamefont {M.}~\bibnamefont {Nakamura}}, \bibinfo
  {author} {\bibfnamefont {K.}~\bibnamefont {Asada}}, \ and\ \bibinfo {author}
  {\bibfnamefont {K.}~\bibnamefont {Hada}},\ }\href@noop {} {\  (\bibinfo
  {year} {2019})},\ \Eprint {http://arxiv.org/abs/1904.05665} {arXiv:1904.05665
  [astro-ph.HE]} \BibitemShut {NoStop}%
\bibitem [{Note3()}]{Note3}%
  \BibitemOpen
  \bibinfo {note} {This is supported by the qualitatively similar dynamical
  processes studied in the cousin model~\cite {Herdeiro:2018wub}.}\BibitemShut
  {Stop}%
\bibitem [{\citenamefont {Tamburini}\ \emph {et~al.}(2019)\citenamefont
  {Tamburini}, \citenamefont {Thidé},\ and\ \citenamefont
  {Della~Valle}}]{Tamburini:2019vrf}%
  \BibitemOpen
  \bibfield  {author} {\bibinfo {author} {\bibfnamefont {F.}~\bibnamefont
  {Tamburini}}, \bibinfo {author} {\bibfnamefont {B.}~\bibnamefont {Thidé}}, \
  and\ \bibinfo {author} {\bibfnamefont {M.}~\bibnamefont {Della~Valle}},\
  }\href@noop {} {\  (\bibinfo {year} {2019})},\ \Eprint
  {http://arxiv.org/abs/1904.07923} {arXiv:1904.07923 [astro-ph.HE]}
  \BibitemShut {NoStop}%
\bibitem [{\citenamefont {{Novikov}}\ and\ \citenamefont
  {{Thorne}}(1973)}]{1973blho.conf..343N}%
  \BibitemOpen
  \bibfield  {author} {\bibinfo {author} {\bibfnamefont {I.~D.}\ \bibnamefont
  {{Novikov}}}\ and\ \bibinfo {author} {\bibfnamefont {K.~S.}\ \bibnamefont
  {{Thorne}}},\ }in\ \href@noop {} {\emph {\bibinfo {booktitle} {Black Holes
  (Les Astres Occlus)}}}\ (\bibinfo {year} {1973})\ pp.\ \bibinfo {pages}
  {343--450}\BibitemShut {NoStop}%
\bibitem [{\citenamefont {Bambi}(2017)}]{Bambi:2017khi}%
  \BibitemOpen
  \bibfield  {author} {\bibinfo {author} {\bibfnamefont {C.}~\bibnamefont
  {Bambi}},\ }\href {\doibase 10.1007/978-981-10-4524-0} {\emph {\bibinfo
  {title} {{Black Holes: A Laboratory for Testing Strong Gravity}}}}\ (\bibinfo
   {publisher} {Springer},\ \bibinfo {year} {2017})\BibitemShut {NoStop}%
\bibitem [{\citenamefont {Doneva}\ \emph {et~al.}(2018)\citenamefont {Doneva},
  \citenamefont {Yazadjiev}, \citenamefont {Stergioulas},\ and\ \citenamefont
  {Kokkotas}}]{Doneva:2018ouu}%
  \BibitemOpen
  \bibfield  {author} {\bibinfo {author} {\bibfnamefont {D.~D.}\ \bibnamefont
  {Doneva}}, \bibinfo {author} {\bibfnamefont {S.~S.}\ \bibnamefont
  {Yazadjiev}}, \bibinfo {author} {\bibfnamefont {N.}~\bibnamefont
  {Stergioulas}}, \ and\ \bibinfo {author} {\bibfnamefont {K.~D.}\ \bibnamefont
  {Kokkotas}},\ }\href {\doibase 10.1103/PhysRevD.98.104039} {\bibfield
  {journal} {\bibinfo  {journal} {Phys. Rev.}\ }\textbf {\bibinfo {volume}
  {D98}},\ \bibinfo {pages} {104039} (\bibinfo {year} {2018})},\ \Eprint
  {http://arxiv.org/abs/1807.05449} {arXiv:1807.05449 [gr-qc]} \BibitemShut
  {NoStop}%
\end{thebibliography}%

 
\end{document}